\documentstyle[preprint,aps,eqsecnum]{revtex}
\tighten
\begin{document}
\title{Relativistic bound-state equations in three dimensions}

\author {D.~R. Phillips and S.~J. Wallace
\footnote{Email: phillips@quark.umd.edu, wallace@quark.umd.edu.}}

\address{Department of Physics and Center for Theoretical Physics,\\
University of Maryland, College Park, MD, 20742-4111}

\maketitle

\begin{abstract}
Firstly, a systematic procedure is derived for obtaining
three-dimensional bound-state equations from four-dimensional ones.
Unlike ``quasi-potential approaches'' this procedure does not involve the
use of delta-function constraints on the relative four-momentum. In the
absence of negative-energy states, the kernels of the three-dimensional
equations derived by this technique may be represented as sums of
time-ordered perturbation theory diagrams. Consequently, such equations
have two major advantages over quasi-potential equations: they may easily
be written down in any Lorentz frame, and they include the
meson-retardation effects present in the original four-dimensional
equation. Secondly, a simple four-dimensional equation with the correct
one-body limit is obtained by a reorganization of the generalized ladder
Bethe-Salpeter kernel. Thirdly, our approach to deriving
three-dimensional equations is applied to this four-dimensional equation,
thus yielding a retarded interaction for use in the 
three-dimensional bound-state equation of Wallace and Mandelzweig.
The resulting three-dimensional equation has the
correct one-body limit and may be systematically improved upon.
The quality of the three-dimensional equation,
and our general technique for deriving such equations,
is then tested by calculating bound-state properties in a scalar field
theory using six different bound-state equations. It is found that
equations obtained using the method espoused here approximate the wave
functions obtained from their parent four-dimensional equations significantly
better than the corresponding quasi-potential equations do.
\end{abstract}

\section {Introduction}

With the advent of experimental facilities such as CEBAF, which are capable of
probing hadronic systems at energies where relativistic effects become 
important, the development of theoretical frameworks in which consistent 
relativistic calculations of few-body hadronic systems can be performed is no
longer merely desirable, it is essential. 

An obvious starting point for such a calculation in a two-body system (such as
the deuteron) is the Bethe-Salpeter equation (BSE) for the four-dimensional, 
covariant, two-to-two amplitude $T$, 
\begin{equation}
T=K + K G_0 T,
\label{eq:BSE}
\end{equation}
where $G_0$ is the free two-particle propagator, which in our convention is
\begin{equation}
G_0(p_1',p_2';p_1,p_2) = i (2 \pi)^8 \delta^{(4)}(p_1' - p_1)
\delta^{(4)}(p_2' - p_2) d_1(p_1) d_2 (p_2),
\end{equation}
with
\begin{equation}
d_i(p_i)=\frac{\Lambda_i^+({\bf p}_i)}{p^0_i - \epsilon_i ({\bf p}_i) + i \eta}
- \frac{\Lambda_i^-({\bf p}_i)}{p^0_i + \epsilon_i ({\bf p}_i) - i \eta},
\label{eq:di}
\end{equation}
where
\begin{eqnarray}
\epsilon_i({\bf p}_i)&=&\sqrt{{\bf p}_i^2 + m_i^2},\\
\Lambda^{\pm}_i({\bf p}_i)&=&\left \{ 
\begin{array}{ll}
\frac{1}{2 \epsilon_i({\bf p}_i)}, & \mbox{for spin-zero particles,}\\
\frac{\pm \epsilon_i ({\bf p}_i) \gamma^0 - {\bf \gamma}_i\cdot {\bf p}_i+m_i}
{2 \epsilon_i({\bf p}_i)}, & \mbox{for spin-half particles,}
\end{array}
\right.
\end{eqnarray}
$\eta$ is a positive infinitessimal 
and $K$ is the Bethe-Salpeter kernel~\cite{BS51,GL50,Na50,Sc51A,Sc51B}.  
In principle $K$ should include all two-particle irreducible two-to-two
Feynman graphs.
Solution of (\ref{eq:BSE}) 
with the full two-particle irreducible kernel is impractical 
and usually resort is made to the ladder
approximation~\cite{LWT67,It67,FT75,FT77}. 
Some undesirable features attend this approximation (see~\cite{Na69}
for a full discussion), not the least of which is the fact that 
the ladder BSE does not have the correct one-body limit~\cite{Gr82}. 
(An equation is said to have the correct one-body limit if, when the
mass of one particle is taken to infinity, the equation reduces to the 
Klein-Gordon or Dirac equation for the light particle 
moving in the static field of the now infinitely massive source.)  
By contrast, 
the full BSE (\ref{eq:BSE}) does possess the correct one-body limit.

Three-dimensional quasi-potential equations (QPEs)
are realized by rewriting the Bethe-Salpeter equation for
the two-particle amplitude (\ref{eq:BSE}) as a pair of coupled
equations:
\begin{eqnarray}
T&=&U + U g T, \label{eq:QPE}\\
U&=&K + K (G_0 - g) U, \label{eq:Ueqn}
\end{eqnarray}
where the propagator $g$ is arbitrary. If $g$ is chosen to contain a 
one-dimensional delta function constraining the relative four-momentum, then 
Eq.~(\ref{eq:QPE}) becomes a three-dimensional equation for the amplitude $T$.
Examples of this approach include the equations of 
Blankenbecler-Sugar~\cite{BbS66} and Gross~\cite{Gr82,Gr69}. 
In principle the driving term $U$ should be determined by solving the 
four-dimensional integral equation (\ref{eq:Ueqn}). However,  
given the difficulty of solving such equations, 
and our lack of knowledge about the best
form of $K$ for hadronic physics, usually $U$ is chosen to be a
one-boson exchange interaction:
\begin{equation}
U=V_{OBE}.
\end{equation}
The t-matrix then obeys the three-dimensional integral equation
\begin{equation}
T=V_{OBE} + V_{OBE} ~g T.
\end{equation}
A reasonable description of few-hadron systems is obtained by
fitting coupling
constants and some mass parameters of $V_{OBE}$ to the
nucleon-nucleon scattering 
data. (Examples of this approach include, but are by no means limited to,
Refs.~\cite{Gr92,HT94,vO95}.)

The QPE formalism of Gross obeys the one-body limit~\cite{Gr82}.  Another 
variant of the quasi-potential approach which respects the one-body
limit was derived by Wallace and Mandelzweig in
References~\cite{MW87,Wa88,WM89}.
We provide a
generalization of that formalism in this paper.  
As originally derived, the QPE of Wallace and Mandelzweig contains, 
for the nucleon-nucleon system, a
one-boson exchange potential which is instant in the center-of-mass frame. 
This involves the use of
a constraint on the relative four-momentum, $p$, of the form 
\begin{equation}
p \cdot P=0 ,
\label{eq:constraint}
\end{equation}
where $P$ is the
total two-body four-momentum.  In the center-of-mass frame, the 
equation for the deuteron vertex function takes the form:
\begin{equation}
\Gamma_{WM}=V_{\mbox{inst}} G_{ET} \Gamma_{WM},
\label{eq:WDinst}
\end{equation}
where $G_{ET}$ refers to the Wallace-Mandelzweig choice for the propagator $g$.
The formalism is covariant since the constraint (\ref{eq:constraint}) 
is expressed covariantly and the 
pair of equations (\ref{eq:WDinst}) and (\ref{eq:Ueqn}) is equivalent to the
bound-state BSE:
\begin{equation}
\Gamma=K G_0 \Gamma.
\end{equation}
Eq.~(\ref{eq:WDinst}) for the cm frame deuteron vertex function
has been used in recent work by Devine and Wallace~\cite{WD94,WD95}. 
This formalism may be developed
in a manifestly covariant fashion following the technique of 
Fuda~\cite{Fu90A}.
This involves writing all four-vectors in terms of their components parallel 
and perpendicular to the total four-momentum:
\begin{equation}
p=p_{\parallel} \hat{P} + p_{\perp}
\end{equation}
where $p_{\parallel}=p \cdot \hat{P}$, with $\hat{P} = P /
\sqrt{P^2}$.  

A fundamental flaw exists in quasipotential formalisms that are based on a 
form for $g$ which contains a delta function. It is generally 
impossible to systematically correct the lowest-order
approximation by use of Eq.~(\ref{eq:Ueqn}) because unphysical
singularities arise. In two-body hadronic systems, such as the deuteron,
calculations may still be pursued using equations such as (\ref{eq:WDinst})
which are themselves free of these singularities.
On the other hand, in any electromagnetic 
reaction where the photon momentum $q$ is non-zero, 
the constraint (\ref{eq:constraint})
cannot hold for the initial and final-state relative four-momenta 
in the cm frame~\cite{WD94}.
Therefore a boosted vertex function is required for electromagnetic matrix
element calculations, or, for that matter, for any QPE calculation in a
three-body problem.
A boost equation for the quasipotential
may be deduced from Eq.~(\ref{eq:Ueqn}), but unphysical
singularities arise in the boosted interaction~\cite{WD94,WD95}.
These are removed from the theory if the full result for $U$, as 
defined by (\ref{eq:Ueqn}), is used in the quasi-potential equation,
but no truncation of $U$ at any finite order in $G-g$ is free of singularities.

Therefore, in this paper we seek a general procedure for the reduction
of four-dimensional equations which {\em does not involve the use of delta
functions}. We develop a procedure by which a 
four-dimensional equation may be approximated 
by a three-dimensional equation and the interaction in that equation 
improved systematically.  
The technique is presented
in Section~\ref{sec-Section2}. It has a close connection to the work of 
Klein~\cite{Kl53,Kl54,KM58,KL74}
on three-dimensional reductions of four-dimensional equations, and to 
standard time-ordered perturbation theory, as discussed in 
Section~\ref{sec-Section3}.

In general the interaction in the three-dimensional equations we
discuss is not covariant. But, since the application of our 
delta-function-free reduction technique to infinite order produces an equation
equivalent to the original four-dimensional equation,
the sum of all terms in the three-dimensional formalism must produce
covariant results. In nuclear physics it is known that 
contributions which are of higher-order in the coupling 
are of increasingly shorter range.  Such short-range
contributions to hadronic interactions must always be treated in an
essentially phenomenological manner.
Therefore, it is expected that a truncation of the interaction at some finite 
order in the coupling
will be useful for applications. It should always be appropriate
to absorb the
effects of the neglected higher-order graphs into phenomenological parameters,
thus ameliorating the non-covariance of the theory.  
In particular, we show in Section~\ref{sec-Section3} that the leading-order 
boost corrections
to the interaction obtained by Forest, Pandharipande and Friar~\cite{Fo95} 
are indeed contained within our truncated interaction.  

In Section~\ref{sec-Section4} we show that the crossed-box graph may be
approximately rewritten as an iterate of the ladder kernel:
\begin{equation}
K_X^{(4)} \approx K^{(2)} G_C K^{(2)},
\end{equation}
where the form of $G_C$ is derived in Section~\ref{sec-Section4}. 
In particular, this approximation is exact in the high-energy (eikonal) and
one-body limits, and may be systematically improved upon in other regimes.
This leads us to propose the four-dimensional integral equation:
\begin{equation}
\Gamma=K^{(2)} (G_0 + G_C) \Gamma.
\label{eq:4DWM}
\end{equation}
This equation has the correct
one-body limit and results from an approximate resummation of the BSE kernel 
$K$. It reduces to the Wallace-Mandelzweig equation 
(\ref{eq:WDinst}) if the dependence of $K^{(2)}$ on the time component 
of the relative four-momentum in the cm frame is neglected.
A three-dimensional reduction of Eq.~(\ref{eq:4DWM}) using the
method developed in Section~\ref{sec-Section2} 
provides a systematic way to include retardation effects, and so improve
on Eq.~(\ref{eq:WDinst}). This reduction to
a three-dimensional equation is performed in Section~\ref{sec-Section5}.
The dynamical boost of the three-dimensional interaction present in the 
equation is realized through simple dependence of the interaction
on the total three momentum and energy of the system, thus eliminating the
boost problem of Eq.~(\ref{eq:WDinst}).
 
In Section~\ref{sec-Section6} the predictions of the
three-dimensional integral equation obtained in Section~\ref{sec-Section5} 
are compared to those of five
other bound-state equations: the ladder BSE and Eq.~(\ref{eq:4DWM}), 
both of which
are, of course, four-dimensional equations, and the three-dimensional equations
of Salpeter, Klein and Wallace \& Mandelzweig. 

The results of Sections \ref{sec-Section3}--\ref{sec-Section5} are  
presented concurrently for a scalar and a spinor field theory.
In particular, in the scalar case we use the $\phi^2 \sigma$
field theory, defined by the Lagrangian
\begin{equation}
{\cal L}=\frac{1}{2}(\partial_\mu \phi_1 \partial^\mu \phi_1
-  m_1^2 \phi_1^2 + \partial_\mu \phi_2 \partial^\mu \phi_2
-  m_2^2 \phi_2^2 + \partial_\mu \sigma
\partial^\mu \sigma - \mu^2 \sigma^2) - 
g_1 m_1 \phi_1^2 \sigma - g_2 m_2 \phi_2^2 \sigma.
\label{eq:Slag}
\end{equation}
Coupling terms include mass factors so that the couplings $g_1$ and $g_2$ are 
dimensionless, thus ensuring that the limits 
$m_1 \rightarrow \infty$  and $m_2 \rightarrow \infty$ correspond to the 
appropriate one-body limits. It is this field theory in which the numerical 
calculations of Section~\ref{sec-Section6} are performed.
In the spinor case the Lagrangian is
\begin{equation}
{\cal L}=\bar{\psi}_1 (i \gamma_\mu \partial^\mu - m_1) \psi_1
+ \bar{\psi}_2 (i \gamma_\mu \partial^\mu - m_2) \psi_2
- g_1 \bar{\psi}_1 \sigma \psi_1 - g_2 \bar{\psi}_2 \sigma \psi_2.
\end{equation}
The arguments of Sections \ref{sec-Section3}--\ref{sec-Section5} are, in fact,
quite general, and, with small modifications, also apply to field theories
involving other types of particles.

\section{A systematic procedure for the removal of the relative-energy 
degree of freedom from a bound-state equation} 

\label{sec-Section2}

Consider the Bethe-Salpeter equation for the bound-state vertex function
$\Gamma$, i.e.,  
\begin{equation}
\Gamma(p;P)  = \int \frac{d^4p'}{(2 \pi )^4} K(p,p';P) G_0(p';P)\Gamma (p';P),
\label{eq:BSBSE}
\end{equation}
Here $P$ is the total two-body
four-momentum, and $p$ and $p'$ are the relative four-momenta.
This equation is completely general, and applies to the bound-state vertex
function in any field theory.
Throughout this paper we find it convenient to omit the explicit integration 
from such equations, abbreviating them as follows
\begin{equation}
\Gamma =  K G_0 \Gamma .  
\end{equation} 

Suppose that the driving term of this equation, $K$, is separated into two
pieces:
\begin{equation}
K=K_1 + K_2,
\end{equation}
where $K_1$ is three-dimensional in the sense that
it does not depend on the zeroth component of the relative four-momenta 
$p$ and $p'$.  For instance, if 
(\ref{eq:BSBSE}) was being solved in the two-body center-of-mass frame and $K$
was a one-boson-exchange kernel then $K_1$ could be chosen to be the usual 
static one-boson-exchange kernel. Note that even though 
the whole driving term $K$ is Lorentz covariant, because the restriction
on $K_1$ is frame dependent, the pieces $K_1$ and $K_2$ are not.
The splitting of $K$ leads to the following coupled equations for $\Gamma$:
\begin{eqnarray}
\Gamma &=& \Gamma _1 + \Gamma _2 ,\\
\Gamma _1 &=&  K_1 G_0 ( \Gamma _1 + \Gamma _2) \label{eq:G1.1},\\
\Gamma _2 &=&  K_2 G_0 ( \Gamma _1 + \Gamma _2 ).
\end{eqnarray}
Both these equations are four-dimensional,  
however, $\Gamma_1$ has no dependence on the zeroth component of
the relative four-momentum because of the defining condition of $K_1$. 
Meanwhile, the second equation may be formally solved to obtain:
\begin{eqnarray}
\Gamma _2 &=& \left[ 1 -  K_2 G_0 \right]^{-1}  K_2 G_0 \Gamma _1 
\nonumber \\
&=& - \Gamma _1 + \left[ 1 -  K_2 G_0 \right]^{-1} \Gamma _1.
\label{eq:Gamma2}
\end{eqnarray}
This last result may then be substituted into Eq.~(\ref{eq:G1.1}) to
yield
\begin{equation}
\Gamma_1 = K_1 {\cal G} \Gamma _1,
\label{eq:G1.2}
\end{equation}
where
\begin{equation}
{\cal G}      =  G_0 +  G_0 K_2 {\cal G}.
\label{eq:GG}
\end{equation}
Eq.~(\ref{eq:G1.2}) then becomes a
three-dimensional equation because the implied integrations over
time-components of momenta only affect ${\cal G}$,  i.e., it reduces to
\begin{equation}
\Gamma _1 = K_1 \langle {\cal G} \rangle \Gamma _1,
\label{eq:RTPT}
\end{equation}
where
\begin{equation}
\langle {\cal G} \rangle\equiv \int \frac {dp_0' \, dp_0}{(2 \pi)^2} 
\, {\cal G}(p_\mu',p_\mu;P).
\label{eq:REint}
\end{equation}
Here we have used the same implied integration notation for the 
three-dimensional integral equation (\ref{eq:RTPT}) as for a four-dimensional
equation. This practice continues below and the context should make it clear
whether the equation in question is a three or four-dimensional one.

Eqs.~(\ref{eq:RTPT}) and (\ref{eq:GG}) are exactly equivalent 
to the original
BSE (\ref{eq:BSBSE}). In order to reconstruct the Bethe-Salpeter amplitude
$\Gamma$ from the non-covariant three-dimensional
$\Gamma_1$ one must use Eq.(\ref{eq:Gamma2}) 
rearranged into the form
\begin{equation}
\Gamma=\left[ 1 -  K_2 G_0 \right]^{-1} \Gamma _1.
\end{equation}

The Green's function, $\langle {\cal G} \rangle$, defined by 
Eq.~(\ref{eq:REint}) is, in fact,
the Fourier transform of the corresponding equal-time co-ordinate
space Green's function 
${\cal G}(t',{\bf x}_1',{\bf x}_2';t,{\bf x}_1,{\bf x}_2)$.
This may be seen by defining
\begin{equation}
{\cal G}(t',{\bf x}_1',{\bf x}_2';t,{\bf x}_1,{\bf x}_2)=
\delta(x_1^{0 \prime} - x_2^{0 \prime}) \delta(x_1^0 - x_2^0) 
g(x_1',x_2';x_1,x_2),
\end{equation}
where $g$ is the appropriate four-dimensional two-body Green's function,
then inserting the integral representation of the delta functions and
taking the Fourier transform~\cite{BK93B,LT63}.

If $K_1$ is now chosen to be a cm-frame instant
interaction, $K^{\mbox{inst}}$,
then (\ref{eq:GG}) may be taken to define ${\cal G}$ order by order
in $K-K^{\mbox{inst}}$. This provides a systematic way to calculate 
retardation corrections
to the Salpeter equation~\cite{Sa52}
\begin{equation}
\Gamma_1 = K^{\mbox{inst}} \langle G_0 \rangle \Gamma _1,
\label{eq:saleq}
\end{equation}
which results from taking ${\cal G}$ at zeroth order in $K-K^{\mbox{inst}}$.
In particular, at first order in $K-K^{\mbox{inst}}$ we find that:
\begin{equation}
\langle {\cal G}^{(1)} \rangle=\langle G_0 \rangle + 
 \langle G_0 K G_0 \rangle -\langle G_0 \rangle K^{\mbox{inst}}
\langle G_0 \rangle.
\end{equation}
The use of this three-dimensional propagator in place of $\langle G_0 \rangle$ 
in (\ref{eq:RTPT}) would therefore
incorporate leading-order retardation corrections 
in the three-dimensional equation.

There is, however, a general way to account for all 
the relative-energy integrations in the modified Green's function 
$\langle {\cal G} \rangle$. Recall that the choice of $K_1$ is  
subject only to the constraint that it should not depend
on the zeroth component of the relative four-momentum in, say, the cm frame.
Consider Eq.~(\ref{eq:GG}) rewritten as
\begin{equation}
\langle {\cal G}\rangle  = \langle G_0 \rangle 
+  \langle G_0 K {\cal G} \rangle  - \langle G_0 \rangle 
 K_1 \langle {\cal G}\rangle,
\label{eq:GG1} 
\end{equation}
and choose $K_1$ such that $\langle {\cal G} \rangle = \langle
G_0 \rangle $.  This provides the defining condition,
\begin{equation}
K_1 \equiv \langle G_0 \rangle ^{-1}  \langle G_0 K {\cal G}
\rangle \langle G_0 \rangle ^{-1} .
\label{eq:K1}  
\end{equation} 
Since this choice of $K_1$ means that $\langle {\cal G} \rangle = \langle
G_0 \rangle$ the full dynamics reduces to a three-dimensional 
integral equation with the free propagator $\langle G_0 \rangle$, i.e.,
\begin{equation}
\Gamma _1 = K_1 \langle G_0 \rangle \Gamma _1,
\label{eq:3Deqn}
\end{equation}
and all the complexities of the relative-energy integrations are
transferred to 
the interaction, $K_1$.

It follows that $K_1$ is the two-particle irreducible (2PI) interaction,
where two-particle irreducibility is defined with respect to 
the three-dimensional propagator $\langle G_0 \rangle$.
To show this formally, rearrange $\langle G_0 K {\cal G} \rangle$ as follows,
\begin{eqnarray}  
\langle G_0 K {\cal G} \rangle 
&=& \langle G_0 K G_0 (1 -  K G_0)^{-1} \rangle  
 + \langle G_0 K G_0 \left\{ [1 -  (K - K_1)G_0 ]^{-1}
- [1 -  K G_0 ]^{-1} \right\} \rangle \nonumber \\
 &=& \langle G_0 T G_0 \rangle - \langle G_0 T G_0 \rangle K_1
\langle {\cal G} \rangle,
\label{eq:rearrange} 
\end{eqnarray}  
where $ T = K ( 1 -  G_0 K) ^{-1} $ is the Bethe-Salpeter
t-matrix defined by Eq.~(\ref{eq:BSE}).  Use of Eq.~(\ref{eq:rearrange}) in 
Eq.~(\ref{eq:K1}) produces 
\begin{equation} 
T_1 = K_1 + T_1 \langle G_0 \rangle K_1,
\end{equation} 
where $T_1 = \langle G_0 \rangle ^{-1} \langle G_0 T G_0 \rangle
\langle G_0 \rangle ^{-1} $ is the three-dimensional t-matrix corresponding
to the Bethe-Salpeter t-matrix $T$. Thus $K_1$ is, indeed, the irreducible
interaction which produces the t-matrix $T_1$ using the
propagator $\langle G_0 \rangle $. We therefore expect that the rules for
the construction of $K_1$ will be akin to those for the two-particle
irreducible interaction in time-ordered perturbation theory. 

\section {Connection to time-ordered perturbation theory and the 
work of Klein}

\label{sec-Section3}

The connection to time-ordered perturbation theory (TOPT)
emerges for a simplified
dynamics in which only positive-energy states are kept in
$G_0$. For the case of a field theory of nucleons and pions it has been
shown that the Green's function $\langle G_0 + G_0 T G_0 \rangle$
obeys the rules of TOPT in the no-anti-nucleon 
case~\cite{BK93B,Ph95}. The same derivations suffice to prove the result
for the scalar field theories defined in the Introduction.
It therefore immediately follows that $K_1$ is 
the 2PI TOPT amplitude in the case that only
positive-energy states are kept in $G_0$.

We now provide a specific example of the derivation of the time-ordered 
perturbation theory amplitude by this route. Firstly, it is true in general
that, if in Eq.~(\ref{eq:K1}) ${\cal G}$ is replaced by $G_0$, then the 
three-dimensional kernel becomes
\begin{equation}
K_1=\langle G_0 \rangle^{-1}  \langle G_0 K G_0 \rangle
\langle G_0 \rangle^{-1}.
\label{eq:K12}
\end{equation}
We note that this formula was given by Klein in
his work on deriving three-dimensional scattering equations from 
four-dimensional ones~\cite{Kl53,Kl54,KM58}. Equations~(\ref{eq:K1}) and 
(\ref{eq:GG}) may thus be thought of as providing
a generalization of the formalism of Klein. 
Secondly, suppose that $K$ is
expanded according to the power of the coupling constant in each of its 
contributions, i.e.,
\begin{equation}
K=\sum_{i=1}^{\infty} K^{(2i)},
\label{eq:Kexp}
\end{equation}
and for the moment only the second-order, or ladder, contribution to
$K$ is kept, i.e., (\ref{eq:K12}) is rewritten as:
\begin{equation}
K^{(2)}_1 = \langle G_0 \rangle ^{-1} \langle G_0 K^{(2)} G_0
\rangle \langle G_0 \rangle ^{-1} .
\label{eq:K13}
\end{equation}
We now calculate this amplitude in the absence of negative-energy states.
First, the propagators are split according to Eq.~(\ref{eq:di}), 
and only the positive-energy pieces retained. Second, the inverse Fourier
representations of all quantities are inserted, so that the time-ordering
can be clearly elucidated. Third,
a change of variables to time differences is made. Fourth, the relative-energy
integrations are performed, generating delta functions on some of the 
time differences. Last, the integrals over these time differences are
performed. 
This procedure allows the calculation of $K_1^{(2)}$ in any frame. 
Indeed, it is completely general, and can be used to calculate $K_1$ no
matter what $K$ is chosen.

If the 
total four-momentum of the two-body system in the frame of interest 
is written 
$P=(E,{\bf P})$, then it is seen that $K_1^{(2)}$ takes the standard form
of TOPT:
\begin{equation}
K_1^{(2)} ({\bf p}',{\bf p};E,{\bf P})= \frac{g_1 g_2}{2 \omega} {\cal M}
\left(\frac{1}{E^+ - \epsilon _1 - \epsilon ' _2 - \omega}
+  \frac{1}{E^+ - \epsilon _2 - \epsilon ' _1 - \omega} \right),
\label{eq:TOPTOPE}
\end{equation}
where $E^+=E + i\eta$,
$\epsilon _1 = \epsilon_1({\bf p}_1)$, 
$\epsilon _2 = \epsilon_2({\bf p}_2)$,
$\omega =\sqrt{\mu^2 + ({\bf p} - {\bf p}')^2 }$, and $\epsilon _1 '$ 
($\epsilon_2 '$) has the argument ${\bf p}_1'$ (${\bf p}_2'$) 
in place of ${\bf p}_1$ (${\bf p}_2$). Here, 
$m_1$ and $m_2$ are the the masses of the two interacting
particles, $\mu$ is the mass of the exchanged meson and the factor 
${\cal M}$ is defined by:
\begin{equation}
{\cal M}=\left \{ 
\begin{array}{ll}
4 m_1 m_2, & \mbox{for spin-zero particles,}\\
1,         & \mbox{for spin-half particles.}
\end{array}
\right.
\label{eq:M}
\end{equation}
Note that to extract the mass of the bound-state,
$M$, one must set $E=\sqrt{M^2 + {\bf P}^2}$.
The total and relative three-momenta
are related to the individual particle momenta in the usual way:
\begin{eqnarray}
{\bf p}_1 + {\bf p}_2&=&{\bf p}_1' + {\bf p}_2'={\bf P};\\
\nu_2 {\bf p}_1 - \nu_1 {\bf p}_2&=&{\bf p}; \, \,
\nu_2 {\bf p}_1' - \nu_1 {\bf p}_2'={\bf p}';
\end{eqnarray}
where $\nu_1$ and $\nu_2$ are any two real numbers which obey 
$\nu_1 + \nu_2=1$. 
Figure \ref{fig-fig1} shows the time-ordered diagrams corresponding to 
Eq.~(\ref{eq:TOPTOPE}).

The bound-state equation corresponding to this interaction is then:
\begin{equation}
\langle G_0 \rangle^{-1} ({\bf p}';E,{\bf P}) \psi({\bf p}';E,{\bf P})
=\int \frac{d^3p}{(2 \pi)^3} 
K_1^{(2)}({\bf p}',{\bf p};E,{\bf P}) \psi({\bf p};E,{\bf P}),
\label{eq:BSE1}
\end{equation}
where we have omitted the factor of 
$(2 \pi)^3 \delta^{(3)}(p'-p)$ from $\langle G_0 \rangle$, which
integrates out of the equation trivially. Here:
\begin{equation}
\langle G_0 \rangle({\bf p};E,{\bf P})=
\frac{\Lambda_1^+ \Lambda_2^+}{E + i\eta - \epsilon_1 - \epsilon_2},
\end{equation}
where $\Lambda_1^+=\Lambda_1^+({\bf p}_1)$ and 
$\Lambda_2^+=\Lambda_2^+({\bf p}_2)$. Now suppose that we have a
set of free single-particle positive-energy momentum-space wave-functions, 
$u$, for both
the spin-half and scalar theories. In the spin-half case these will
be the spinors, and their normalization may be chosen such that:
\begin{equation}
\Lambda^+ ({\bf p})=u({\bf p}) \bar{u}({\bf p})
\label{eq:projector}
\end{equation}
(where spin indices have been suppressed).
In the scalar case Eq.~(\ref{eq:projector}) may 
also be enforced, since the
$u$'s may be chosen to be merely numerical factors, i.e., we may define
\begin{equation}
u({\bf p}) = \bar{u}({\bf p}) \equiv \frac{1}{\sqrt{2 \epsilon({\bf p})}}.
\end{equation}
Expanding:
\begin{equation}
\psi({\bf p};E,{\bf P})=u_1({\bf p}_1) u_2 ({\bf p}_2) \phi({\bf p};E,{\bf P}),
\end{equation}
(with an implicit summation on spin indices in the spinor case)
and manipulating Eq.~(\ref{eq:BSE1}) leads to:
\begin{equation}
(E - \epsilon_1' - \epsilon_2') \phi({\bf p}';E,{\bf P})=
\int \frac{d^3 p}{(2 \pi)^3} {\tilde K}_1^{(2)}({\bf p}',{\bf p};E,{\bf P})
\phi({\bf p};E,{\bf P}),
\label{eq:BSE2}
\end{equation}
with:
\begin{equation}
{\tilde K}_1^{(2)}({\bf p}',{\bf p};E,{\bf P})={\bar u}_1({\bf p}_1') 
{\bar u}_2({\bf p}_2') K_1^{(2)}({\bf p}',{\bf p};E,{\bf P})
u_1({\bf p}_1) u_2({\bf p}_2).
\end{equation}

Note that the TOPT interaction which appears in the bound-state equation
(\ref{eq:BSE2})
changes with total three-momentum of the
system.  This provides the dynamical boost of the interaction
and it is straightforward to show that it incorporates the
perturbative boost correction discussed in
Ref.~\cite{Fo95}. That is, in the equal-mass case, to second order in ${\bf P}$
we have (neglecting ${\bf p}^2/m$ and ${\bf p}'^2/m$ terms):
\begin{equation}
{\tilde K}_1^{(2)}({\bf q}; {\bf P}) = {\tilde K}_1^{(2)}({\bf q};{\bf 0}) 
- \frac{{\bf P}^2}{4m^2} {\tilde K}_1^{(2)}({\bf q};{\bf 0}) - 
\frac{1}{8m^2} {\bf P} \cdot {\bf q} {\bf P} \cdot {\bf \nabla}_{\bf q}
{\tilde K}_1^{(2)}({\bf q};{\bf 0}),
\label{eq:Fboost}
\end{equation}
where we have written the interaction as a function of the momentum of the 
exchanged meson, ${\bf q}={\bf p}' - {\bf p}$. 

If it is true, as argued 
in Ref.~\cite{Fo95}, that this leading-order boost
correction is sufficiently accurate for studies of 
three-nucleon
systems, then the boost effects incorporated in the TOPT result 
(\ref{eq:TOPTOPE})
should be more than adequate. Indeed, we believe such a boost to be preferable
to that of Eq.~(\ref{eq:Fboost}),
since it involves operators which are bounded as ${\bf P} \rightarrow \infty$,
whereas, at large values of ${\bf P}$ the perturbative boost correction 
(\ref{eq:Fboost}) diverges.

The identification of the kernel $K_1$ with the usual time-ordered 
perturbation theory interaction is only correct in the absence of 
negative-energy states. 
With the full four-dimensional propagator, the form of $K_1^{(2)}$  
obtained from Eq.~(\ref{eq:K13}) in the manner described above 
Eq.~(\ref{eq:TOPTOPE}) is:
\begin{eqnarray}
&&\langle G_0 \rangle({\bf p}';E,{\bf P})
K_1^{(2)} ({\bf p}', {\bf p};E,{\bf P})
\langle G_0 \rangle({\bf p};E,{\bf P})=\nonumber\\
&& \quad \frac{g_1 g_2 {\cal M}}{2 \omega} \left[
\left (\frac{{\Lambda_1^+}' {\Lambda_2^+}'}{E-\epsilon_1'-\epsilon_2'}  + 
\frac{{\Lambda_1^-}' {\Lambda_2^+}'}{-\epsilon_1' - \epsilon_1 - \omega} \right)
\frac{1}{E - \epsilon_1 - \epsilon_2' - \omega}
\left(\frac{\Lambda_1^+ \Lambda_2^+}{E-\epsilon_1-\epsilon_2} + 
\frac{\Lambda_1^+ \Lambda_2^-}{-\epsilon_2 - \epsilon_2' - \omega}\right) 
\right.\nonumber\\
&&\quad + \left(\frac{{\Lambda_1^+}'{\Lambda_2^+}'}{E-\epsilon_1'-\epsilon_2'}
+\frac{{\Lambda_1^+}'{\Lambda_2^-}'}{-\epsilon_2 - \epsilon_2' - \omega}\right)
\frac{1}{E - \epsilon_1' - \epsilon_2 - \omega}
\left (\frac{\Lambda_1^+ \Lambda_2^+}{E-\epsilon_1-\epsilon_2}  +
\frac{\Lambda_1^- \Lambda_2^+}{-\epsilon_1' - \epsilon_1 - \omega} \right) 
\nonumber\\
&& \quad + \frac{{\Lambda_1^-}' {\Lambda_2^-}'}{-E-\epsilon_1'-\epsilon_2'}
\left(\frac{1}{-\epsilon_1' - \epsilon_1 - \omega} + 
\frac{1}{-\epsilon_2' - \epsilon_2 - \omega}\right)
\frac{\Lambda_1^+ \Lambda_2^+}{E-\epsilon_1-\epsilon_2}\nonumber\\
&& \quad +\left(\frac{{\Lambda_1^-}'{\Lambda_2^-}'}{-E-\epsilon_1'-\epsilon_2'}
+ \frac{{\Lambda_1^+}' {\Lambda_2^-}'}{-\epsilon_1 -\epsilon_1'-\omega}\right) 
\frac{1}{-E-\epsilon_1-\epsilon_2'-\omega}
\left(\frac{\Lambda_1^- \Lambda_2^-}{-E-\epsilon_1-\epsilon_2}
+ \frac{\Lambda_1^- \Lambda_2^+}{-\epsilon_2 - \epsilon_2' - \omega}\right)
\nonumber\\
&&\quad +\left(\frac{{\Lambda_1^-}' {\Lambda_2^-}'}{-E-\epsilon_1'-\epsilon_2'}
+ \frac{{\Lambda_1^-}' {\Lambda_2^+}'}{-\epsilon_2 -\epsilon_2'-\omega}\right)
\frac{1}{-E-\epsilon_1'-\epsilon_2-\omega}
\left(\frac{\Lambda_1^- \Lambda_2^-}{-E-\epsilon_1-\epsilon_2}
+ \frac{\Lambda_1^+ \Lambda_2^-}{-\epsilon_1 - \epsilon_1' - \omega}\right)
\nonumber\\
&& \quad \left. +\frac{{\Lambda_1^+}'{\Lambda_2^+}'}{E-\epsilon_1'-\epsilon_2'}
\left(\frac{1}{-\epsilon_1' - \epsilon_1 - \omega} +
\frac{1}{-\epsilon_2' - \epsilon_2 - \omega}\right)
\frac{{\Lambda_1^-} {\Lambda_2^-}}{-E-\epsilon_1-\epsilon_2}\right],
\label{eq:Kleinpot}
\end{eqnarray}
where $\Lambda_i^{\pm}=\Lambda_i^{\pm} ({\bf p}_i)$ and 
${\Lambda_i^{\pm}}'=\Lambda_i^{\pm} ({\bf p}_i')$. 
Note that here, and throughout the rest of this paper, the $i \eta$ 
prescriptions can be generated by understanding the masses $m_1$, $m_2$, and
$\mu$ to have a small negative imaginary part.

Now the three-dimensional propagator is
\begin{equation}
\langle G_0 \rangle({\bf p};E,{\bf P})=
\frac{\Lambda_1^+ \Lambda_2^+}{E + i\eta - \epsilon_1 - \epsilon_2} + 
\frac{\Lambda_1^- \Lambda_2^-}{- E + i\eta - \epsilon_1 - \epsilon_2}
\label{eq:G0ave}
\end{equation}
(where once again we have omitted the momentum-space delta-function
for notational simplicity),
which is the Salpeter propagator~\cite{Sa52}. If the negative-energy piece
is removed, this becomes the 
Blankenbecler-Sugar~\cite{BbS66} or Logunov-Tavkhelidze~\cite{LT63} propagator.
Note that in the spin-half case this propagator does not 
have a unique inverse, consequently in that case 
the derivation of the bound-state equation
analogous to Eq.~(\ref{eq:BSE2}) usually assumes that the $+-$ and $-+$ 
pieces of $\langle G_0 \rangle^{-1}$ are zero.

The graphs corresponding to the terms in Eq.~(\ref{eq:Kleinpot}) are shown in 
Figure~\ref{fig-fig2}. 
Graphs (a)-(d) are generated by the first line of Eq.~(\ref{eq:Kleinpot}),
graphs (e)-(h) by the second line, and graphs (i) and (j) by the
third line. The fourth to sixth lines of (\ref{eq:Kleinpot}) generate the same 
ten graphs, but
with particles and anti-particles interchanged. Note that the graphs
drawn in Figure~\ref{fig-fig2} do not represent actual physical processes,
but rather contributions to the ``potential'' $K_1^{(2)}$.
As observed by Klein, if negative-energy particles are present, the rules for 
the construction of this interaction differ from 
those of standard TOPT~\cite{Kl53,Kl54}.

In general,
it can be shown that the rules for the construction of 
the full two-body equal-time Green's function, 
$\langle G_0 + G_0 T G_0 \rangle$,
for the case of positive-energy initial and final-state particles,
are, in either of the two field theories given in the Introduction:
\begin{enumerate}
\item Draw all topologically distinct, two-to-two graphs,
which do not contain vacuum-vacuum subdiagrams, remembering
that different time-orderings contribute to different graphs. 

\item Construct the expression for each individual graph exactly as in
TOPT, save that:
\begin{enumerate}
\item All one-particle energies are to be relativistic.

\item If in some intermediate state of the graph both the initial-state 
particles are present with exactly the same momenta as in their initial state, 
i.e., their state is unaltered from the beginning of the graph, 
then in the TOPT denominator corresponding to that state the quantity 
$\epsilon_1 + \epsilon_2$ must be replaced by $E$.

\item Similarly, if both final-state particles are present with exactly
the same momenta as in their final state, i.e., they have  undergone their 
last interaction but are still present in some
intermediate state, then in the TOPT denominator 
corresponding to that state the quantity $\epsilon_1' + \epsilon_2'$
must be replaced by $E$.
\end{enumerate}
\end{enumerate}
If both these last two conditions are satisfied then the denominator
for the relevant state is the
TOPT denominator for that state, but with $\epsilon_1' + \epsilon_2' + 
\epsilon_1 + \epsilon_2$ replaced by $2E$. 
Note that these are not the rules for 
the amplitude obtained from the Ladder BSE. The Green's function
defined by these rules
is the $++ \rightarrow ++$ piece of the {\em full} 
two-body equal-time 
Green's function of the field theory.

The substitution of these $E$s in the intermediate-state denominators can
be shown to be due to a crucial difference between the way initial and 
final-states are treated in the calculation of $\langle G_0 T G_0 \rangle$ 
and in TOPT. In TOPT the initial and final times in the diagram are
taken to minus and plus
infinity respectively, thus guaranteeing that no interaction takes
place before the beginning (or after the end) 
of the propagation of the two-particle
state. However, when $\langle G_0 T G_0 \rangle$ is calculated in our work the
initial and final times are kept finite. Consequently the first (last) event
in the diagram need not be the creation of the $\phi_1 \phi_2$ or $N_1 N_2$
pair, it may
be the creation out of (destruction into) the vacuum of a $\phi \bar{\phi} 
\sigma$ or $N \bar{N} \sigma$ state. 
(See for instance the second graph in Figure~\ref{fig-fig2}.) 
The possibility of such an
event happening before or after the $\phi_1 \phi_2$ or $N_1 N_2$ propagation 
in some contributions to $\langle G_0 T G_0 \rangle$
leads to denominators which differ from the TOPT ones 
in those pieces of $\langle G_0 T G_0 \rangle$. 
It follows that, if the pieces of the Hamiltonian 
of the theory which couple $\phi \bar{\phi} \sigma$ or $N \bar{N} \sigma$
states directly to the vacuum are removed from the Hamiltonian,
then the modification of the TOPT denominators 
is not necessary, and the 
rules for calculation of the two-particle to two-particle part of the
Green's function $\langle G_0 T G_0 \rangle$
become exactly those for the calculation of the TOPT
Green's function. 
This equivalence of $\langle G_0 T G_0 \rangle$ and the time-ordered
perturbation theory Green's function in the absence of such terms 
from the Hamiltonian was demonstrated for a field theory of 
nucleons and pions by Kvinikhidze and Blankleider~\cite{BK93B}.

As discussed in Section~\ref{sec-Section2}, the use of a 
static interaction in the Bethe-Salpeter
equation leads to the Salpeter equation (\ref{eq:saleq}). This equation
is often used for the interactions of two relativistic particles. It is
straightforward to show that the static limit of the potential given by 
Eq.~(\ref{eq:Kleinpot}) is the usual instant interaction 
in the positive-energy sector.
On the other hand, Eq.~(\ref{eq:Kleinpot}) predicts that
the coupling between the $++$ and $--$ states 
is suppressed by retardation effects which cause it to
be a factor $\omega/m$ (in the equal-mass case) smaller than the rest of the 
static interaction. This factor is zero in the static limit.
Thus, for a causal interaction, the 
correct static limit of the equation (\ref{eq:3Deqn}) with the interaction
(\ref{eq:Kleinpot}) is the Breit equation, i.e.,
Eq.~(\ref{eq:saleq}) with the $--$ states omitted from the calculation.

In order to recover the results of Klein for two-meson exchange interactions,
and so make connection with the recent work on two-pion exchange of 
Rijken and Stoks~\cite{Ri91,RS95} and Lahiff and Afnan~\cite{LA96}, 
we evaluate 
${\cal G}$ to first order in $K - K_1$ in
Eq.~(\ref{eq:K1}).  This produces
\begin{equation}
K_1=\langle G_0 \rangle^{-1} (\langle G_0 K G_0 \rangle +  \langle G_0 K G_0
(K - K_1) G_0 \rangle) \langle G_0 \rangle^{-1}.
\label{eq:K1fourth}
\end{equation}
Now expanding $K$ as in Eq.~(\ref{eq:Kexp}) and dropping terms
of higher than fourth order in the coupling constant leads to
\begin{equation}
K_1^{(4)}=\langle G_0 \rangle^{-1}  \biggl( \langle G_0 K^{(4)} G_0 \rangle + 
 \langle G_0 K^{(2)} G_0 K^{(2)} G_0 \rangle \biggr) \langle G_0 \rangle^{-1}
- K_1^{(2)} \langle G_0 \rangle K_1^{(2)}.
\label{eq:K14}
\end{equation}
When applied to particle-particle scattering Eq.~(\ref{eq:K14}) gives 
exactly the same results obtained by Klein, except that where
Klein removed the iterated second-order three-dimensional
interaction by hand, here the formalism provides  
the subtraction naturally.

The procedure developed in Section~\ref{sec-Section2} thus provides the
generalization of 
Klein's method.  The interaction and vertex function defined 
by Eqs.~(\ref{eq:K1}), (\ref{eq:GG}), and 
(\ref{eq:3Deqn}) have the full
relative-energy dynamics of the Bethe-Salpeter equation (\ref{eq:BSBSE})
included in them.
 
\section {A simple four-dimensional equation with the correct one-body limit}

\label{sec-Section4}

Now let us return to the four-dimensional BSE (\ref{eq:BSE}). 
The simplest BSE with the correct one-body limit is
\begin{equation}
\Gamma=K_X G_0 \Gamma,
\label{eq:3.1}
\end{equation}
where $K_X$ is the sum of all 2PI ladder and crossed-ladder two-to-two 
graphs~\cite{Gr82}.
The bound-state masses predicted by such an equation have recently been
obtained by Monte Carlo integration in the Feynman-Schwinger representation
of both scalar $\phi^3$ field theory and scalar QED by Nieuwenhuis, Tjon
and Simonov~\cite{TS93,Ne94,Ne95,NT96,NT95}. 
In general though, the kernel $K_X$ is too 
complicated for 
(\ref{eq:3.1}) to be solved by standard means. Therefore, in this section we 
shall derive a four-dimensional equation 
which has a simple kernel, the appropriate meson-production thresholds and the 
correct one-body and high-energy (or eikonal) limits.
A form of the following derivation appeared in~\cite{Wa88}.

Although it is impossible to rewrite a crossed graph exactly as
an iterate of
the ladder kernel $K^{(2)}$, the leading contributions of these graphs to the 
high-energy and one-body limits are iterative.  
This can be easily shown in the case of the fourth-order crossed-box
graph. 
In both of the field theories defined in the Introduction, this graph 
corresponds to an expression:
\begin{eqnarray}
&& K_X^{(4)}(k_1',k_2';k_1,k_2)=\nonumber\\
&& \quad i g_1^2 g_2^2 {\cal M}^2 \int \frac{d^4 p_2}{(2 \pi)^4} 
\frac{1}{(k_2' - p_2)^2 - \mu^2} d_1(P-p_2) d_2(k_2 + k_2' - p_2)
\frac{1}{(p_2 - k_2)^2 - \mu^2}.
\label{eq:3.6}
\end{eqnarray}
where the lines have been assigned the momenta shown in Figure
\ref{fig-Xedbox}, and $P$ is the total momentum, which is conserved:
\begin{equation}
P=k_1 + k_2=k_1' + k_2'.
\end{equation}
The propagators $d_i$ were given in Eq.~(\ref{eq:di}), and the factor 
${\cal M}$ was defined in Eq.~(\ref{eq:M}).

Following Wallace and Mandelzweig~\cite{Wa88,WM89} we define, 
for any four-vector $q$, quantities $q_{\parallel}$ 
and ${q_{\perp}}_\mu$, 
as follows:
\begin{equation}
q_\mu=q_{\parallel} \hat{P}_\mu + {q_{\perp}}_\mu,
\end{equation}
where $\hat{P}$ is the unit four-vector in the direction of $P$ and
\begin{equation}
q_{\parallel}=q \cdot \hat{P}. 
\end{equation}
Consequently in Eq.~(\ref{eq:3.6}) the argument of the function $d_2$ may be
rewritten:
\begin{equation}
({k_2}_{\parallel} + {k_2'}_{\parallel} - {p_2}_{\parallel})
\hat{P} + {k_2}_{\perp} + {k_2'}_{\perp} - {p_2}_{\perp}.
\end{equation}

Now suppose that the momentum of particle 2 is large.  This may
occur because 
$m_2 \gg m_1$ (one-body limit) or because particle 2 has very high energy
(eikonal limit).  In either case its 
intermediate and final-state momenta will be largely 
unaffected by the presence of particle 1, and so we may approximate
the perpendicular components as unchanging,
\begin{equation}
{k_2}_{\perp} + {k_2'}_{\perp} \approx 2 {p_2}_{\perp}.
\label{eq:eikapprox}
\end{equation}
Indeed, making the replacement (\ref{eq:eikapprox}) in
(\ref{eq:3.6}) will not affect the value of $K_X^{(4)}$ in the limit $m_2 
\rightarrow \infty$.

This argument shows that $K_X^{(4)}$ may be approximately rewritten as:
\begin{eqnarray}
&& K_X^{(4)}(k_2',k_2;P) \approx \nonumber\\
&& \quad i \int \frac{d^4 p_2}{(2 \pi)^4} 
K^{(2)}(k_2' - p_2) d_1(P-p_2) d_2(({k_2}_{\parallel} + {k_2'}_{\parallel}
- {p_2}_{\parallel}) \hat{P} + {p_2}_{\perp}) K^{(2)}(p_2-k_2),
\end{eqnarray}
where $K^{(2)}$ is the ladder BSE driving term:
\begin{equation}
K^{(2)}(q)={\cal M} \frac{g_1 g_2}{q^2 - \mu^2 + i\eta}.
\end{equation}
In operator notation:
\begin{equation}
K_X^{(4)} \approx K^{(2)} G_C K^{(2)},
\label{eq:KXit}
\end{equation}
with
\begin{equation}
G_C(p',p;E_1,E_2)=i (2 \pi)^4 \delta^{(4)} (p'-p)
d_1(p)d_2((E_2 - E_1 + p_\parallel)\hat{P} - p_{\perp}).
\label{eq:GC1}
\end{equation}
Here $E_1$ and $E_2$ are defined via:
\begin{equation}
2 E_2={k_2}_{\parallel} + {k_2'}_{\parallel}; \quad
2 E_1={k_1}_{\parallel} + {k_1'}_{\parallel};
\end{equation}
and so in the cm frame $E = E_1 + E_2$. 
The propagator $G_C$ defined by (\ref{eq:GC1}) therefore depends on the 
parallel components of the external momenta. Thus the use of
operator notation in Eq.~(\ref{eq:KXit}) is not strictly correct.
In Eq.~(\ref{eq:newGC}) we redefine $G_C$ in order to remove this
dependence on external momenta. However, 
that change in $G_C$ modifies the analytic structure of the amplitude defined
by the corresponding integral equation.  Therefore, for the present
we persist with $G_C$ defined by (\ref{eq:GC1}), and use an improper
operator notation. 
Note that if the particles are on-shell in their initial and final
states then:
\begin{equation}
E_1=E_1^{\mbox{on}} \equiv \frac{E^2 + m_1^2 - m_2^2}{2 E}; \quad
E_2=E_2^{\mbox{on}} \equiv \frac{E^2 + m_2^2 - m_1^2}{2 E}.
\end{equation}
These arguments show that Eq.~(\ref{eq:KXit}) will be exact
in the infinite-mass and high-energy limits, thus demonstrating that the pieces
of $K_X^{(4)}$ which survive in these two limits may, indeed, be written as
iterates of $K^{(2)}$.

Now suppose that $K_X$ is written as
\begin{equation}
K_X=V + V G_C K_X.
\label{eq:inteqKX}
\end{equation}
In principle this is always possible, as (\ref{eq:inteqKX}) may be taken as 
a definition of $V$.
At second order in the coupling we clearly have:
\begin{equation}
V^{(2)}=K_X^{(2)} \equiv K^{(2)},
\end{equation}
while the above argument shows that with this $V^{(2)}$
\begin{equation}
K_X^{(4)}=V^{(2)} G_C V^{(2)},
\end{equation} 
in the high-energy and infinite-mass limits. Thus, a reasonable choice
for $V$ is $V=K^{(2)}$. Equation (\ref{eq:inteqKX}) then defines corrections to
this choice.

Once this $V$ is chosen, 
Eq.~(\ref{eq:inteqKX}) and the BSE (\ref{eq:3.1}) may be combined to
yield an ``improved'' ladder BSE, which, in the two-body cm frame, 
after a change of variables to total and relative four-momenta, 
takes the form
\begin{eqnarray}
\Gamma(p_0',{\bf p}';s)&=&i \int \frac{d^4 p}{(2 \pi)^4}
K^{(2)}(p'-p) d_1(E_1^{\mbox{on}}+p_0,{\bf p})\nonumber\\
&& \left[d_2(E_2^{\mbox{on}}-p_0,-{\bf p}) + 
d_2(E_2^{\mbox{on}} - p_0' + p_0,-{\bf p}) \right]  \Gamma(p_0,{\bf p};s).
\label{eq:3.19}
\end{eqnarray}

We stress that what has been done here is to take certain pieces of the 
Bethe-Salpeter kernel $K_X$ and rewrite them in the form $K^{(2)} G_C K^{(2)}$,
$K^{(2)} G_C K^{(2)} G_C K^{(2)}$, etc. Consequently, Eq.~(\ref{eq:3.19}) 
is equivalent to a Bethe-Salpeter equation in which graphs other than one-meson
exchange are approximately
included in the kernel. Thus we expect that the solution of this
equation  may provide a better description of the dynamics of two-particle 
systems than the ladder BSE amplitude. 

This will be especially true for systems with one particle much
heavier than the other. In these systems the one-body limit constitutes
an important piece of the dynamics. Unlike the ladder BSE, Eq.~(\ref{eq:3.19})
has the correct one-body limit.
This may be shown as follows. For the on-shell vertex function $p_0'=0$.
If $p_0'=0$ the sum of the two particle 2 propagators, multiplied by ${\cal M}$
is, in the spin-zero case:
$$
{\cal M} \left[d_2(E_2^{\mbox {on}}-p_0,-{\bf p}) + 
d_2(E_2^{\mbox {on}} + p_0,-{\bf p})\right]=
\frac{4 m_1 m_2}{(E_2^{\mbox {on}} - p_0)^2 - \epsilon_2^2 + i\eta} + 
\frac{4 m_1 m_2}{(E_2^{\mbox {on}} + p_0)^2 - \epsilon_2^2 + i\eta},
$$
while for the spin-half case the same combination becomes:
$$
\frac{(E_2^{\mbox {on}} - p_0) \gamma^0 + {\bf \gamma}_2 
\cdot {\bf p} + m_2}{(E_2^{\mbox {on}} - p_0)^2 - \epsilon_2^2 + i\eta} + 
\frac{(E_2^{\mbox {on}} + p_0) \gamma^0 + {\bf \gamma}_2 \cdot {\bf p}
 + m_2}{(E_2^{\mbox {on}} + p_0)^2 - \epsilon_2^2 + i \eta}.
$$
In the $m_2 \rightarrow \infty$ limit, 
$E_2^{\mbox {on}} \rightarrow \epsilon_2 \rightarrow m_2$, 
and so this expression reduces to
\begin{equation}
\left[ \frac{1}{p_0 + i \eta} 
- \frac{1}{p_0 - i\eta}\right] 2 m_1 = - 2 \pi i \delta (p^0) 2 m_1,
\end{equation}
for the spin-zero case and
\begin{equation}
- 2 \pi i \delta (p^0) \Lambda_{2 \infty}^+,
\end{equation}
for the spin-half case, where
\begin{equation}
\Lambda_{2 \infty}^+=\frac{1 + \gamma_0}{2},
\end{equation}
is the positive-energy projection operator for an infinitely massive particle 
two.
Thus, in the infinite $m_2$ limit 
Eq.~(\ref{eq:3.19}) yields the Klein-Gordon or Dirac equation for 
the wave function of particle 1 moving in the static $\sigma$ field
generated by particle 2
\begin{equation}
\frac{E^2 - m_1^2 - {\bf p}'^2}{2 m_1} \psi({\bf p}')=
-\int \frac{d^3 p}{(2 \pi)^3}  
\frac{g_1 g_2}{({\bf p}' - {\bf p})^2 - \mu^2} \psi({\bf p}),
\end{equation}
or
\begin{equation}
(E \gamma^0 - {\bf p}' \cdot {\bf \gamma} - m_1) \psi({\bf p}')=
-\int \frac{d^3 p}{(2 \pi)^3}
\frac{g_1 g_2}{({\bf p}' - {\bf p})^2 - \mu^2} \psi({\bf p}),
\end{equation}
where
\begin{equation}
\psi({\bf p})=\frac{1}{E^2 - m_1^2 - {\bf p}^2}
\Gamma(0,{\bf p};E)
\end{equation} 
or 
\begin{equation}
\psi({\bf p})=\frac{1}{E \gamma^0 - {\bf p} \cdot {\bf \gamma} - m_1}
\Lambda_{2 \infty}^+ \Gamma(0,{\bf p};E),
\end{equation}
and $E=E_1^{\mbox{on}}$.

Equation (\ref{eq:3.19}) also has the appropriate meson-production thresholds,
and therefore could be used as a basis for four-dimensional calculations
of two-body bound-state properties.  
However, the equation cannot be written in the form
\begin{equation}
\Gamma= K^{(2)} G \Gamma,
\label{eq:3.19b}
\end{equation}
and consequently the method of Section~\ref{sec-Section2} cannot
be applied to it. Therefore, we now seek an approximate version of 
Eq.~(\ref{eq:3.19}) which can be written in the form (\ref{eq:3.19b}).

In the on-shell vertex function $p_0'=0$. 
In what follows we use $p_0'=0$ in the
integrand also when the
amplitude is 
not on-shell. This approximation provides a 
four-dimensional version of the
three-dimensional Wallace-Mandelzweig equation obtained in 
Ref.~\cite{MW87,Wa88,WM89}. In the center-of-mass frame it is
\begin{eqnarray}
\Gamma(p';E_1^{\mbox {on}},E_2^{\mbox {on}})&=&i \int 
\frac{d^4 p}{(2 \pi)^4} 
K^{(2)}(p'-p) d_1(E_1^{\mbox {on}}+p_0,{\bf p}) \nonumber\\
&& \left[d_2(E_2^{\mbox {on}}-p_0,-{\bf p}) + 
d_2(E_2^{\mbox {on}} + p_0,-{\bf p}) \right] 
\Gamma(p;E_1^{\mbox {on}},E_2^{\mbox {on}}).
\label{eq:3.20}
\end{eqnarray}
Redefining $G_C$  to be, in the cm frame,
\begin{equation}
G_C(p',p;E_1^{\mbox {on}},E_2^{\mbox {on}})=i (2 \pi)^4 
\delta^{(4)}(p'-p)
d_1(E_1^{\mbox {on}}+p_0,{\bf p})
d_2(E_2^{\mbox {on}} + p_0,-{\bf p}),
\label{eq:newGC}
\end{equation}
allows Eq.~(\ref{eq:3.20}) to be written:
\begin{equation}
\Gamma=K^{(2)} (G_0 + G_C) \Gamma.
\label{eq:modified4Dapp}
\end{equation}

The following points are worth noting:
\begin{enumerate}
\item Because the new choice for $G_C$ does not affect the on-shell
fourth-order piece of the amplitude Eq.~(\ref{eq:KXit}) is
still exactly true on-shell in the one-body and eikonal limits.

\item In the limit $m_2 \rightarrow \infty$ Eq.~(\ref{eq:3.20}) has the 
correct one-body limit, as can be seen by a similar argument to that given for 
Eq.~(\ref{eq:3.19}).

\item If a pinch analysis of the singularities of the amplitude $T$ defined
by the scattering equation corresponding to (\ref{eq:3.20}) is performed then 
it is found that at order $g^4$ there is a production cut in 
the $p_0'$ plane of the half-off-shell amplitude which extends upwards from
\begin{equation}
p_0'=m_2 + \mu - E_2^{\mbox {on}},
\end{equation}
and lies infinitessimally below the real axis.
In the on-shell amplitude at sixth order this cut appears in the 
integrand and overlaps with the pole of the 
Green's function at $p_0'=E_2^{\mbox {on}} - \epsilon_2 ({\bf q})$, 
so producing a cut in the
$E_2^{\mbox {on}}$ plane of $T$ which extends upwards from
\begin{equation}
E_2^{\mbox {on}}=m_2 + \frac{\mu}{2}.
\label{eq:E2thresh}
\end{equation}
This is not a particle-production threshold which exists in the 
full Bethe-Salpeter scattering amplitude. Hence 
its  existence in the amplitude $T$ defined by the scattering equation 
corresponding to Eq.~(\ref{eq:3.20}) must be
regarded as a deficiency of that equation.
It arises because the method used to derive Eq.~(\ref{eq:3.20}) is only 
guaranteed to produce the correct cut-structure for the second and 
fourth-order on-shell amplitudes. 
It should be noted that the vertex function given by
Eq.~(\ref{eq:3.19}) does not contain this anomalous threshold.

The value of $E$ at which the threshold (\ref{eq:E2thresh}) occurs
in the on-shell amplitude is
$E=m_2 + \frac{\mu}{2}+\sqrt{m_1^2 + m_2 \mu + \frac{\mu^2}{4}}$.
In fact, since Equation~(\ref{eq:3.19}) was designed for use in the 
$m_2 > m_1$ regime this is actually above $E=m_1 + m_2 + \mu$.
Therefore the theory has the correct threshold structure 

for all energies $E$ such that $m_2 - m_1 < E < m_1 + m_2 + \mu$.
Furthermore, if $m_1=m_2$ then the threshold (\ref{eq:E2thresh}) 
actually lies at $E=m_1 + m_2 + \mu$, 
which is where the first production threshold of the crossed-box graph 
should be. Consequently, we think of this cut as representing the usual 
single-pion-production threshold of the crossed-box graph, but somewhat 
displaced if $m_2 > m_1$.

\item This equation was derived for particles of different mass. If we desire 
an equation for identical particles a symmetric form of the propagator $G_C$
must be used. (We shall return to this point in the next section.)

\item The above derivation could equally well be pursued for the exchange of
vector particles. However, in, for instance, scalar QED, account must be taken 
of the seagull graphs. Furthermore, in the case of spinor QED more care 
must be taken, since an
additional piece of the interaction may be generated when one attempts to write
$X \approx V G_C V$. (The interested reader may consult \cite{WM89} 
for details on these points.)
\end{enumerate}

\section{A three-dimensional equation with retardations and the correct
one-body limit}

\label{sec-Section5}

We now apply the relative-energy integration method of 
Section~\ref{sec-Section2} to the four-dimensional
equation~(\ref{eq:3.20}). By so doing we first recover Wallace 
and Mandelzweig's
original quasi-potential equation, and then calculate a 
three-dimensional kernel which includes first-order retardation corrections 
to this result.

Observe that once $G_C$ is defined by
Eq.~(\ref{eq:newGC}),
Eqs.~(\ref{eq:inteqKX}) and (\ref{eq:3.1}) are equivalent to:
\begin{equation}
\Gamma=V (G_0 + G_C) \Gamma,
\label{eq:modified4D}
\end{equation}
where $V$ is regarded as being {\em defined} by (\ref{eq:inteqKX}). 
This provides a method for calculating 
corrections to Eq.~(\ref{eq:modified4Dapp}), 
which was obtained by assuming that 
$V=K^{(2)}$.
The result of applying the method
of Section~\ref{sec-Section2} to Eq.~(\ref{eq:modified4D}) is:
\begin{eqnarray}
\Gamma_1&=&V_1 \langle {\cal G} \rangle \Gamma_1, \label{eq:4.1}\\
{\cal G}&=&G_0 + G_C + (G_0 + G_C) (V - V_1) {\cal G}. \label{eq:4.2}
\end{eqnarray}

If $V$ is taken to be $V_{\sigma \mbox{inst}}$, where $V_{\sigma \mbox{inst}}$
is the static one-sigma exchange potential,
\begin{equation}
V_{\sigma \mbox{inst}}({\bf p}',{\bf p})=-\frac{g_1 g_2 {\cal M}}
{({\bf p} - {\bf p}')^2 + \mu^2},
\label{eq:Vsiginst}
\end{equation}
these equations reduce to
\begin{equation}
\Gamma_1=V_{\sigma \mbox{inst}} G_{ET} \Gamma_1 ,
\label{eq:4.3}
\end{equation}
which is a three-dimensional equation with a free two-body propagator that
in the cm frame takes the form:
\begin{eqnarray}
G_{ET}({\bf p};E_1^{\mbox {on}},E_2^{\mbox {on}})&\equiv&
\langle G_0 + G_C \rangle\\
\:\:\: &=& \frac{\Lambda_1^+ \Lambda_2^+}{E_1^{\mbox {on}} 
+ E_2^{\mbox {on}} - \epsilon_1 - \epsilon_2 + i \eta} + 
\frac{\Lambda_1^+ \Lambda_2^-}{E_1^{\mbox {on}} - E_2^{\mbox {on}} 
- \epsilon_1 - \epsilon_2 + i \eta} \nonumber\\
&& \quad + \frac{\Lambda_1^- \Lambda_2^+}{-E_1^{\mbox {on}} + E_2^{\mbox {on}} 
- \epsilon_1 - \epsilon_2 + i \eta} + 
\frac{\Lambda_1^- \Lambda_2^-}{-E_1^{\mbox {on}} 
- E_2^{\mbox {on}} - \epsilon_1 - \epsilon_2 + i \eta},
\label{eq:4.5}
\end{eqnarray}
where once again a factor of $(2 \pi)^3 \delta^{(3)}(p'-p)$ has been
omitted for notational simplicity.
Note that this propagator is equivalent to that derived by Cooper and 
Jennings by a different technique~\cite{CJ89}. Note also that in the case 
$m_1=m_2$ this same propagator is referred to as the equal-time propagator
by Tjon and collaborators~\cite{HT94,Ne95,NT96,TT92,TT93}.

The desired generalization of Eq.~(\ref{eq:4.3}) which incorporates
retardation and boost effects follows from applying the
ideas of Sections~\ref{sec-Section2} and \ref{sec-Section3} to 
Eq.~(\ref{eq:4.2}). 
If $V_1$ is chosen 
such that the last term in Eq.~(\ref{eq:4.2}) is 
zero, the formula
\begin{equation}
V_1=G_{ET}^{-1} \langle (G_0 + G_C) V {\cal G} \rangle G_{ET}^{-1},
\end{equation}
is obtained for the general three-dimensional interaction to be used in
\begin{equation}
\Gamma_1=V_1 G_{ET} \Gamma_1.
\label{eq:REEWM}
\end{equation}
A first step in the inclusion of retardation and boost effects may be
taken by following Klein's work and replacing
$\cal G$ by $G_0 + G_C$.  This yields what we refer to as the
first-order relative-energy integration result:
\begin{eqnarray}
V_1&=&G_{ET}^{-1} A G_{ET}^{-1} \label{eq:4.7}\\
A&=&\langle (G_0 + G_C) V (G_0 + G_C) \rangle.
\label{eq:4.8}
\end{eqnarray}

We now take $V$ equal to $K^{(2)}$ and
calculate $A$ in the center-of-mass frame. In order to do this we first
decompose the cm frame propagators:
\begin{equation}
d_i(p)=\sum_{\rho_i} 
\frac{\Lambda^{\rho_i}_i ({\bf p})}{\rho_i p_0 - \epsilon_i ({\bf p})}; 
\quad i=1,2,
\end{equation}
where $\rho_i$ may take on the values $\pm 1$.
It follows that the amplitude $A$ may be written as
\begin{equation}
A=\sum_{\rho_1 \rho_1' \rho_2 \rho_2'} \Lambda^{\rho_1' \prime}_1 
\Lambda^{\rho_2' \prime}_2
A(\rho_1' \rho_2' \leftarrow \rho_1 \rho_2) \Lambda^{\rho_1}_1 
\Lambda^{\rho_2}_2.
\label{eq:4.11}
\end{equation}
The sixteen contributions $A(\rho_1' \rho_2' \leftarrow \rho_1 \rho_2)$ 
may be calculated by the same method used to obtain the Klein
potential (\ref{eq:Kleinpot}). They are found to 
take on a slightly different form depending on whether $\rho_1' \rho_1$ is 
equal to plus or minus one. In order to simplify the expressions for the
$A$'s we define the following quantities:
\begin{eqnarray}
F&=&\frac{g_1 g_2 {\cal M}}{2 \omega};\\
e_i=\rho_i E_i^{\mbox{on}} - \epsilon_i; && e_i'=\rho_i' E_i^{\mbox{on}} 
- \epsilon_i'; \:\:\: i=1,2.
\end{eqnarray}

If $\rho_1' \rho_1=1$ then:
\begin{eqnarray}
A(\rho_1' \rho_2' \leftarrow \rho_1 \rho_2)({\bf p}',{\bf p};
E_1^{\mbox {on}},E_2^{\mbox {on}})
&=&F \frac{1}{e_1' + e_2'} \frac{1}{e_1 + e_2' - \omega} 
\left[\frac{1}{e_1 + e_2} + \frac{1}{e_2 + e_2' - \omega}\right]\nonumber\\
&+& F \left[\frac{1} {e_1' + e_2'} 
+ \frac{1}{e_2 + e_2' -\omega}\right] 
\frac{1}{e_1' + e_2 - \omega} \frac{1}{e_1 + e_2}.
\label{eq:4.15}
\end{eqnarray}
Conversely, if $\rho_1' \rho_1=-1$ then:
\begin{eqnarray}
A(\rho_1' \rho_2' \leftarrow \rho_1 \rho_2)({\bf p}',{\bf p};
E_1^{\mbox {on}},E_2^{\mbox {on}})
&=& F \frac{1}{e_1 + e_1' - \omega} 
\frac{1}{e_1 + e_2' - \omega} \left[\frac{1}
{e_1 + e_2} + \frac{1}{e_2 + e_2' - \omega}\right]\nonumber\\
&+& F \left[\frac{1} {e_1' + e_2'} + \frac{1}{e_2 + e_2' - \omega}
\right]  \frac{1}{e_1' + e_2 - \omega}
\frac{1}{e_1 + e_1' - \omega}\nonumber\\
&+& F \frac{1}{e_1' + e_2'} \left[\frac{1}{e_1 + e_1' - \omega}  
+ \frac{1}{e_2 + e_2' - \omega} \right] \frac{1}{e_1 + e_2}.
\label{eq:4.16}
\end{eqnarray}

We note the following points about the interaction defined by 
Eqs.~(\ref{eq:4.5}), (\ref{eq:4.7}),  (\ref{eq:4.11}), 
(\ref{eq:4.15}) and (\ref{eq:4.16}).
\begin{enumerate}
\item As expected from the four-dimensional equation (\ref{eq:3.20})
it assumes a static form in the limit $m_2 \rightarrow
\infty$. In other words, it has the correct one-body limit.
(Recall that we assumed that particle 2 was the heavier of the two.)

\item In the static limit of the interaction defined by Eqs.~(\ref{eq:4.15})
and (\ref{eq:4.16}) all couplings to the $--$
states are suppressed. On the other hand, 
the other couplings tend to the
same static limit as the $++ \rightarrow ++$ piece of the interaction. 
Thus, in this limit the 
$++ \rightarrow +-$ coupling, which is due to Z-graphs, is correctly given
by the instantaneous exchange interaction, and only the $--$ states
need be omitted from a calculation. The role of the $+-$ and $--$
states in this limit is therefore consistent
with the one-body limit which motivated Eq.~(\ref{eq:REEWM}).

\item Also as expected from (\ref{eq:3.20}) $A (++ \leftarrow ++)$ has a 
singularity structure which is different to that of the ordinary time-ordered 
perturbation theory amplitude: it contains a 
cut beginning at $E_2 = m_2 + \frac{\mu}{2}$. 
The same comments made about the unusual singularity structure of 
(\ref{eq:3.20}) apply to this cut. 

\item Because we have used a form of $G_C$ derived by considering the
limit $m_2 \rightarrow \infty$, the interaction is {\em not} symmetric
under the interchange of particle 1 and 2 labels.
\end{enumerate}

Points (2) and (3) are not a reflection of the underlying physics
of the meson-exchange process, but rather of the particular iterative 
form we used in our attempt to sum some of the higher-order graphs 
in the kernel of the BSE (\ref{eq:3.1}). 

The choice of $G_C$ used above involves an approximation 
which is appropriate in the regime $m_2 > m_1$. For equally massive
particles a more appropriate choice for the propagator $G_C$ is:
\begin{eqnarray}
G_C(p',p;E_1^{\mbox{on}},E_2^{\mbox{on}})=
i (2 \pi)^4 \delta^{(4)}(p'-p)
\frac{1}{2} \left[d_1(E_1^{\mbox{on}} + p_0,{\bf p}) 
d_2(E_2^{\mbox{on}} + p_0,-{\bf p})\right.\nonumber\\
\left. + d_1(E_1^{\mbox{on}} - p_0,{\bf p})
d_2(E_2^{\mbox{on}} - p_0,-{\bf p})\right].
\end{eqnarray}
Note that this choice is label symmetric, and hence the resulting bound-state
equation may be used for identical particles.
The $G_C$ defined by (\ref{eq:newGC}) did not have this property.

When Eqs.~(\ref{eq:4.7}) and (\ref{eq:4.8}) are applied with this symmetrized 
form of $G_C$ 
the result is a $V_1$ of the form (\ref{eq:4.11}), but with an
$A(\rho_1' \rho_2' \leftarrow \rho_1 \rho_2)$ 
which may be written in any $\rho$-spin channel as:
\begin{eqnarray}
&& A({\bf p}',{\bf p};E_1^{\mbox{on}},E_2^{\mbox{on}})=\nonumber\\
&& \quad \frac{F}{2} \left[\frac{1}{e_1' + e_2'} 
\left(\frac{1}{e_1 + e_2' - \omega}
+ \frac{1}{e_1' + e_2 - \omega} + \frac{1}{e_1 + e_1' - \omega}
+ \frac{1}{e_2 + e_2' - \omega}  \right)
\frac{1}{e_1 + e_2} \right.\nonumber\\
&& \quad \quad +  \frac{1}{e_1' + e_2'} \frac{1}{e_1 + e_2' - \omega} 
\frac{1}{e_2 + e_2' - \omega}
+ \frac{1}{e_1' + e_2'} \frac{1}{e_1' + e_2 - \omega}
\frac{1}{e_1 + e_1' - \omega}\nonumber\\
&& \quad \quad +  \frac{1}{e_1 + e_1' - \omega} \frac{1}{e_1 + e_2' - \omega}
\frac{1}{e_1 + e_2} 
+ \frac{1}{e_2 + e_2' - \omega} \frac{1}{e_1' + e_2 - \omega}
\frac{1}{e_1 + e_2}\nonumber\\
&& \quad \quad + \left. \frac{1}{e_2 + e_2' - \omega} 
\left(\frac{1}{e_1 + e_2' - \omega}
+ \frac{1}{e_1' + e_2 - \omega} \right) 
\frac{1}{e_1 + e_1' - \omega} \right],
\label{eq:V1symm}
\end{eqnarray}
with all symbols defined as above.

\section {Comparison of the bound-state predictions of different three and
four-dimensional integral equations}

\label{sec-Section6}

Having derived the interaction which is to be used in Eq.~(\ref{eq:REEWM}) 
we may now compare the two-body
bound-state properties predicted by this integral equation with those 
properties obtained from other
three and four-dimensional calculations. This is done in the scalar field 
theory defined in the Introduction. The masses $m_1$, $m_2$, and 
$\mu$ are chosen to be $m_1=m_2=m$ and $\mu=0.15 m$. Units are then chosen
so that $\hbar=c=m=1$.

The three-dimensional equations to be considered are:
\begin{enumerate}
\item The Salpeter equation, which may be written:
\begin{equation}
\Gamma_{S}=V_{\sigma \mbox{inst}} \langle G_0 \rangle \Gamma_{S},
\label{eq:SalBS}
\end{equation}
with $V_{\sigma \mbox{inst}}$ given by Eq.~(\ref{eq:Vsiginst}) and 
$\langle G_0 \rangle$ by Eq.~(\ref{eq:G0ave}). Note that in this scalar
field theory Eq.~(\ref{eq:SalBS}) is exactly equivalent to the 
Blankenbecler-Sugar equation obtained from the ladder BSE.

\item The Klein equation for a one-sigma-exchange interaction, which is 
Eq.~(\ref{eq:3Deqn}) with the interaction defined by Eqs.~(\ref{eq:Kleinpot})
and (\ref{eq:G0ave}).

\item The equation proposed by Wallace and Mandelzweig, which we refer to as 
the ET equation:
\begin{equation}
\Gamma_{ET}=V_{\sigma \mbox{inst}} G_{ET} \Gamma_{ET},
\end{equation}
where $G_{ET}$ was given in Eq.~(\ref{eq:4.5}).

\item The retarded ET equation, Eq.~(\ref{eq:4.1}), with the label-symmetric
interaction obtained in the previous section, i.e., with $V_1$ defined by 
(\ref{eq:V1symm}), (\ref{eq:4.11}), (\ref{eq:4.7}) and (\ref{eq:4.5})
\end{enumerate}

The four-dimensional equations these four calculations are to be compared to 
are:
\begin{enumerate}
\item The ladder Bethe-Salpeter equation.

\item Equation~(\ref{eq:3.20}) which, unlike the ladder BSE, has the correct
one-body limit.
\end{enumerate}

We consider the six bound-state equations as falling into two groups: the
first contains the ladder BSE and the two three-dimensional reductions 
thereof: the Salpeter equation and the Klein equation. The
second group contains the four-dimensional Wallace-Mandelzweig equation 
(\ref{eq:3.20}) and the two three-dimensional equations derived from it:
the equal-time equation (\ref{eq:4.3}) 
and the first-order relative-energy integration 
equation (\ref{eq:REEWM}) with
the interaction $V_1$ defined by Eqs.~(\ref{eq:V1symm}),
(\ref{eq:4.11}), (\ref{eq:4.7}) and (\ref{eq:4.5}). Within these two
groups the three equations may then be thought of as being zeroth (instant
potential equations), first (Klein-type) and infinite (four-dimensional 
equations) order relative-energy integration results.

At any given bound-state mass the equation:
\begin{equation}
\Gamma(m)= M(m) \Gamma(m),
\label{eq:4.21}
\end{equation}
whether it be a three or four-dimensional integral equation,
is an eigenvalue problem. The coupling constant $g_1 g_2$ which 
appears in the kernel must be chosen such that $M(m)$ has eigenvalue one. 
If we are searching for the ground-state of the system then we may assume that
$\Gamma$ is an $S$-wave state, thus implying that the function $\Gamma$ has no
angular dependence. The integration over $\hat{p}$ may then be easily 
performed. The resultant one (or two) dimensional integral equation may then be
discretized, reducing the problem to one of solving for the
eigenvalues of a matrix version of $M$. 
Note that in the four-dimensional equations it is necessary to perform
a Wick rotation in the variable $p_0$ before the kernel $M$ 
is discretized, as otherwise the analytic structure in the kernel makes
this procedure numerically unstable.

When this approach is implemented for the equations listed above 
the six bound-state spectra shown in Figure~\ref{fig-bsspectra} are obtained. 
The following observations may immediately be made:
\begin{enumerate}
\item The two instant formalisms (Salpeter and ET) predict more binding at a 
given 
coupling than either their corresponding four-dimensional equation or their 
corresponding Klein-type interaction.

\item The calculations
in the second group always predict deeper binding than the corresponding 
calculations in the first group. In other words, the inclusion of pieces of 
higher-order graphs in the kernel of the integral equation yields deeper 
binding at a given coupling.

\item The first-order relative-energy integration (Klein-type)
interactions give predictions for the
bound-state spectra which are significantly closer to those of the full
four-dimensional calculation than the spectra of the instant equations.
\end{enumerate}

All of these results can be understood on physical grounds. Point 1 arises 
because the inclusion of retardation in the scalar field theory always 
reduces the amount of attraction in the bound-state. An example of this is
provided by one-sigma exchange. When the full retardation corrections
predicted by time-ordered perturbation theory are included
the instant potential (\ref{eq:Vsiginst}) is replaced by
the retarded interaction (\ref{eq:TOPTOPE}) (provided the effects
of negative-energy states are ignored).
If $E < m_1 + m_2$ this will always be a negative number that is smaller in 
magnitude than the instant potential. Meanwhile, the inclusion of pieces
of additional higher-order graphs leads to deeper binding since in the 
$\phi^2 \sigma$ theory {\em all} time-ordered perturbation theory graphs are 
attractive in the bound-state region. Most significant though, in our opinion, 
is point 3, which shows that the first-order relative-energy integration 
(Klein-type) potentials do a 
better job than the instant interactions
of approximating the four-dimensional equation which both are derived from. 

All six of these calculations aim to sum the main contributions 
to the two-to-two amplitude. However, Figure
\ref{fig-bsspectra} shows that they give different results, and so they cannot
all be correct.
The question therefore arises: what is the 
binding energy at a given coupling if the full scattering series is summed? 
A partial answer to this question was recently provided by Nieuwenhuis
and Tjon, who summed the series of all ladder and crossed-ladder 
diagrams using 
Monte Carlo integration techniques in the Feynman-Schwinger representation, 
and so obtained two-body bound-state 
masses in this scalar theory~\cite{TS93,Ne94,Ne95,NT96}. Their results, 
calculated at six 
different coupling strengths, show that the sum of all such graphs predicts 
deeper binding than {\em any} of our six integral equation calculations. In 
particular, the ET appears to do the best job of reproducing the bound-state 
spectrum obtained by Nieuwenhuis and Tjon. 
At first sight this appears to recommend the ET as the best formalism for 
doing three-dimensional calculations. However, two points must be remembered
before a definite conclusion is drawn.

Firstly, the reason that the Nieuwenhuis-Tjon calculation predicts so much more
binding than
the two four-dimensional integral equation calculations performed here
is precisely the fact
discussed in connection with point 2 above: in the $\phi^2 \sigma$ theory 
{\em all} time-ordered 
perturbation theory graphs give
attraction in the bound-state region. Hence the Nieuwenhuis-Tjon calculation, 
which includes
many more such graphs than both the ladder BSE and four-dimensional 
Wallace-Mandelzweig equation 
calculations, {\em must} predict more binding. 
Therefore, we claim that the Salpeter and ET
equations do a ``better'' job of imitating the full Nieuwenhuis-Tjon 
calculation than the ladder
BSE and four-dimensional Wallace-Mandelzweig equation because instant 
calculations such as the Salpeter and ET ones
ignore two competing effects. 
\begin{enumerate}
\item Higher-order graphs are left out of their kernel. If included these
graphs would lead to deeper binding at a given coupling.

\item Exchanged-meson retardation is completely ignored. If included this
retardation would lead to shallower binding at a given coupling.
\end{enumerate}
In other words, we claim that the ET formalism's apparent success is 
due to the
cancellation of these two neglected effects and cannot necessarily be 
interpreted as a signal of that 
approach having the ``right physics'' for the problem.

Secondly, {\em none} of these calculations have been performed in the way that 
calculations in few-body hadronic physics are usually performed. When
such bound-state calculations are done in hadronic physics one usually 
acknowledges that the use of {\em any} integral equation necessarily means the
neglect of some graphs of the theory, and so one regards the couplings that 
appear in that equation as effective. These couplings are then usually 
fixed by fitting to some set of observables, and then certain different
observables are predicted. This makes the connection
to an underlying field theory somewhat more tenuous, but it does allow 
calculation to proceed in situations  where:
\begin{itemize}
\item The value of the coupling which appears in the
``fundamental'' Lagrangian is not known.

\item The problem is highly non-perturbative and consequently {\em any} 
truncation is somewhat suspect.
\end{itemize}

Therefore, to test the procedure of 
Section~\ref{sec-Section2} for deriving three-dimensional integral equations
 in a way consistent with that in
which it would be used in hadronic physics, we fix the mass of the bound-state 
at $M=1.95$ and calculate the wave-functions
\begin{equation}
\psi({\bf p}';E)=g({\bf p}';E)
\Gamma({\bf p}';E),
\end{equation}
for each of the four different three-dimensional calculations. (Here $g$ is the
free two-particle Green's function for the relevant calculation.)
Each three-dimensional wave-function must be normalized according to
\begin{equation}
\int \frac{d^3p' \, d^3 p}{(2 \pi)^6} \psi^*({\bf p}';E) 
\frac{\partial G^{-1}}{\partial E}({\bf p}',{\bf p};E) \psi({\bf p};E)=2E,
\label{eq:3Dnormn}
\end{equation}
where $G$ is the full two-particle Green's function appearing in the 
three-dimensional equation from which $\Gamma$ was obtained. 
These wave-functions are then 
compared to the four-dimensional wave functions, integrated over the zeroth 
component of the relative four-momentum:
\begin{equation}
\psi({\bf p}';E)=\frac{1}{2 \pi} \int dp_0' \, g(p_\mu';E) 
\Gamma(p_\mu';E).
\label{eq:4Dwf}
\end{equation}
Here $g$ is the appropriate free two-particle 
four-dimensional Green's function.
The four-dimensional vertex function used in (\ref{eq:4Dwf}) is 
normalized according to:
\begin{equation}
\int \frac{d^4p' d^4 p}{(2 \pi)^8} \Gamma^*(p_\mu';E) g(p_\mu';E) 
\frac{\partial G^{-1}}{\partial E}(p_\mu',p_\mu;E)g(p_\mu;E)\Gamma(p_\mu;E)=2E,
\label{eq:4Dnormn}
\end{equation}
where $G$ is the full two-particle Green's function appearing in the 
four-dimensional equation from which $\Gamma$ was obtained.
The left-hand sides of Eqs.~(\ref{eq:3Dnormn}) and (\ref{eq:4Dnormn}) may be 
reduced to two (and four)-dimensional integrals using the fact
that the $S$-wave wave-function $\psi$ is angle-independent. Note that in the 
four-dimensional case, Wick rotation in the zeroth component allows 
use of the vertex function on the $p_0$ or $p_0'$ imaginary axis.

When the normalized wave-functions of the equations in group 1 (group 2) 
are compared the results shown in Figure~\ref{fig-wfcomp1} (\ref{fig-wfcomp2})
are obtained. It is immediately seen
that all the wave-functions at this bound-state mass have essentially the same
features. However, in both figures it is clear that the three-dimensional 
equation with the first-order relative-energy integration, or Klein-type, 
interaction reproduces the 
integrated four-dimensional wave function considerably more accurately than
the instant calculation does.
This shows that the procedure given in Section 2 {\em does} provide a way to 
systematically derive three-dimensional equations which are closer to their
four-dimensional counterparts than those which are obtained from merely 
adopting an instant interaction.

\section {Conclusion}

Quasi-potential equations (QPEs), despite their success as a basis for hadronic
phenomenology, cannot be improved upon systematically. This is because whenever
they are used beyond first-order they predict amplitudes with unphysical
singularities. These singularities also appear whenever attempts are made
to boost the interaction appearing in the QPE from
one frame to another. They arise from these equations' use of delta-function 
constraints 
on the relative four-momentum.

Because of this shortcoming of QPEs, in Section~\ref{sec-Section2} of this 
paper we 
sought, and found, a systematic procedure for deriving three-dimensional
equations from four-dimensional ones which does not involve the use of
delta functions. As shown in Section~\ref{sec-Section3}, this procedure
is akin to the work of Klein. It allows the derivation of a three-dimensional
kernel which includes, in a systematic expansion which may be pursued to any
desired accuracy,
the effects of the relative-energy
integration present in the four-dimensional integral equation. 
If the procedure is 
applied to infinite order, a result equivalent to the original four-dimensional
equation is obtained. At any order the kernel derived has a simple
interpretation
in terms of the diagrams of time-ordered perturbation theory, except that
the presence of negative-energy states requires additional rules. 
The resultant 
three-dimensional equations therefore include more of the meson-retardation
effects than quasi-potential equations, and incorporate the effects
of a dynamical boost.

Since the BSE with any kernel which is a finite sum of Feynman graphs 
does not have the correct one-body limit, if the procedure of 
Section~\ref{sec-Section2} is directly applied to any solvable BSE a 
three-dimensional
equation without the correct one-body limit is found. However, as
demonstrated by Wallace~\cite{Wa88}, and recapitulated here in 
Section~\ref{sec-Section4}, a Bethe-Salpeter equation with the correct
one-body limit may be reorganized so that the pieces of the kernel which
contribute at leading order in the one-body limit take on an iterative
form. Hence a four-dimensional equation, Eq.~(\ref{eq:3.19}), 
with the correct one-body limit and meson-production thresholds is derived.
A further approximation leads to a simple four-dimensional equation to which
the procedure of Section~\ref{sec-Section2} may be applied, 
Eq.~(\ref{eq:3.20}). In making that
approximation some of the meson-production thresholds are displaced.
Nevertheless, the
equation still has exactly the correct cut-structure for $m_2 - m_1 < E <
m_1 + m_2 + \mu$.

In Section~\ref{sec-Section5} the procedure of Section~\ref{sec-Section2} was
applied to the four-dimensional equation (\ref{eq:3.20}). The result is a
simple three-dimensional equation with the correct one-body limit which
has straightforward boost properties. The main shortcoming of this equation 
is the fact
that certain thresholds of the original Bethe-Salpeter wave function are 
modified in the three-dimensional wave function. However, this modification
is due to the approximation in moving from (\ref{eq:3.19}) to (\ref{eq:3.20}),
rather than to any deficiencies of our procedure for deriving
three-dimensional equations. It could be systematically corrected for,
at the price of complicating the three-dimensional kernel.

Finally, in Section~\ref{sec-Section6} we compared and contrasted the 
bound-state properties predicted by six different bound-state equations 
in a scalar field theory. A first group of equations contained
the ladder Bethe-Salpeter equation
and two three-dimensional equations based on it: the Salpeter equation, and
the equation obtained by applying our method to first order---which we 
referred to as the Klein equation, since the formula obtained for the
three-dimensional interaction appears in Klein's work. In the second group
were Eq.~(\ref{eq:3.20}), the so-called equal-time
equation and the equation  derived in Section~\ref{sec-Section6}. It
was found that in each group of three equations the equation
derived by the Klein-like delta-function-free reduction technique of 
Section~\ref{sec-Section2} approximated its parent four-dimensional equation
better than the corresponding instant equation did. However, the bound-state
spectra from the instant ET equation lies closest to the bound-state spectrum
obtained from the sum of ladder and crossed-ladder graphs, as calculated by
Nieuwenhuis and Tjon~\cite{TS93,Ne94,Ne95,NT96}. This can be understood on 
physical grounds and, we argued, is not necessarily a recommendation
for the use of the ET equation in physical systems. Indeed, when the
wave functions of a two-body system in a scalar field theory are examined 
it is seen that the equation derived in Section~\ref{sec-Section5} does
a much better job of reproducing the integrated four-dimensional
wave function than the ET equation wave function does. 

The results of Section~\ref{sec-Section6} are in harmony with
recent results for phase shifts in scalar-scalar scattering obtained
by Lahiff and Afnan~\cite{LA96}. They found that the Klein method
reproduces
the phase shifts obtained from the corresponding Bethe-Salpeter equation much 
better than
a Blankenbecler-Sugar calculation with an instant interaction.

These formal and numerical results indicate that the procedure
of Section~\ref{sec-Section2} is successful in its goal of providing a 
way to systematically obtain three-dimensional equations from 
four-dimensional ones without the use of delta functions. In particular,
when applied to the four-dimensional equation (\ref{eq:3.20}) this
procedure yields a three-dimensional equation which, unlike the corresponding
quasi-potential equation (\ref{eq:4.3}), has a well-defined boost.
This would seem to make such an 
equation a good starting point for calculations in few-hadron systems.

\acknowledgements{S.~J.~W. acknowledges a conversation in which Mr. Paul 
Dulany suggested that delta-function constraints were unnecessary in the
three-body problem and a three-dimensional reduction of the Bethe-Salpeter
amplitude could be effected by integrating out relative energies.  This
idea stimulated the present work on the two-body problem.
D.~R.~P. thanks Coen van Antwerpen for his help in 
writing the code used to solve the Bethe-Salpeter equation.
We are grateful to the U.S. Department of Energy for its support 
under contract no. DE-FG02-93ER-40762.}

\begin{figure}
\caption{The two graphs which contribute to the second-order
three-dimensional kernel, if only positive-energy intermediate states
are included in the calculation.}
\label{fig-fig1}
\end{figure}

\begin{figure}
\caption{The ten graphs which contribute to the second-order
three-dimensional kernel if propagation into negative-energy 
states is allowed. Note that these graphs can be interpreted as
applying to positive or negative-energy particles. In fact, 
each graph shown contributes twice: the second time with 
positive-energy particles going forward in time interchanged with
negative-energy particles going backward. The dotted lines represent
the initial and final times in each graph. Observe that in all but
graphs (a) and (e) some interactions take place outside the interval
$[t_i,t_f]$.}
\label{fig-fig2}
\end{figure}

\begin{figure}
\caption{The crossed-box graph, showing the momentum labels used
in the text.}
\label{fig-Xedbox}
\end{figure}

\begin{figure}
\caption{A plot of bound-state mass ($\frac{M}{m}$) versus coupling strength 
($\frac{g_1 g_2}{\pi}$) for the seven different calculations discussed in
the text. The points with error bars are taken from {\protect \cite{Ne95}} and
represent the FSR calculation of Nieuwenhuis and Tjon. The other calculations
were all performed using three or four-dimensional integral equations, with
the legend as indicated in the figure.}
\label{fig-bsspectra}
\end{figure}

\begin{figure}
\caption{A plot of normalized wave-functions $k \psi(k)$ versus momentum $k$ 
(in units such that $m=1$), for a bound-state with $M=1.95 m$. The different 
wave functions shown are the integrated ladder BSE wave function (solid line); 
the result of the 
Klein potential calculation (long dashed line) and the Salpeter equation
wave function (dotted line).}
\label{fig-wfcomp1}
\end{figure}

\begin{figure}
\caption{A plot of normalized wave-functions $k \psi(k)$ versus momentum $k$
(in units such that $m=1$), for a bound-state with $M=1.95 m$. The different
wave functions shown are the integrated four-dimensional
Wallace-Mandelzweig equation wave function (thick solid line);
the wave-function resulting from the first-order relative-energy integration 
method applied to that
equation (dash-dotted line) and the ET wave function (short dashed line).}
\label{fig-wfcomp2}
\end{figure}

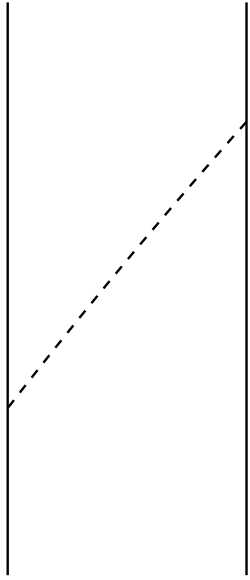
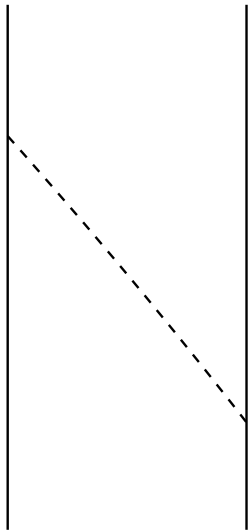
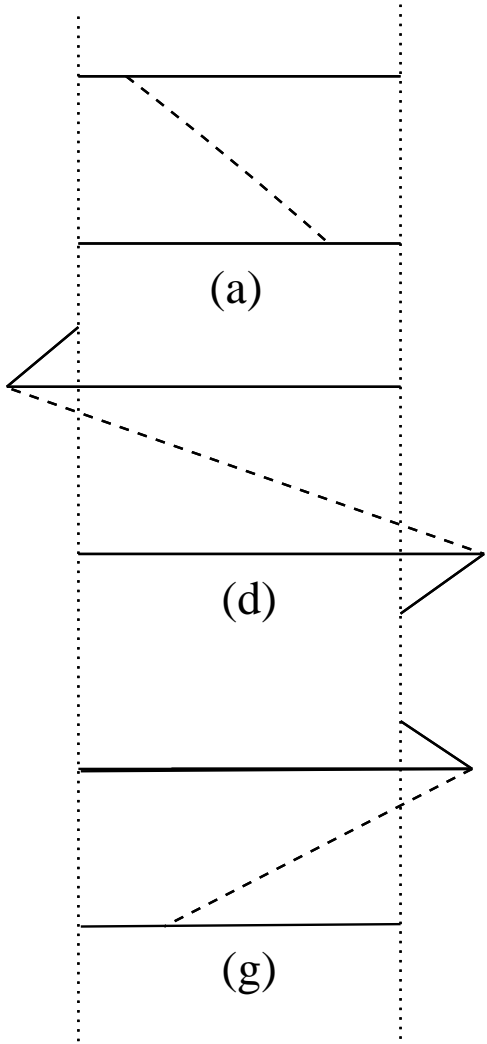
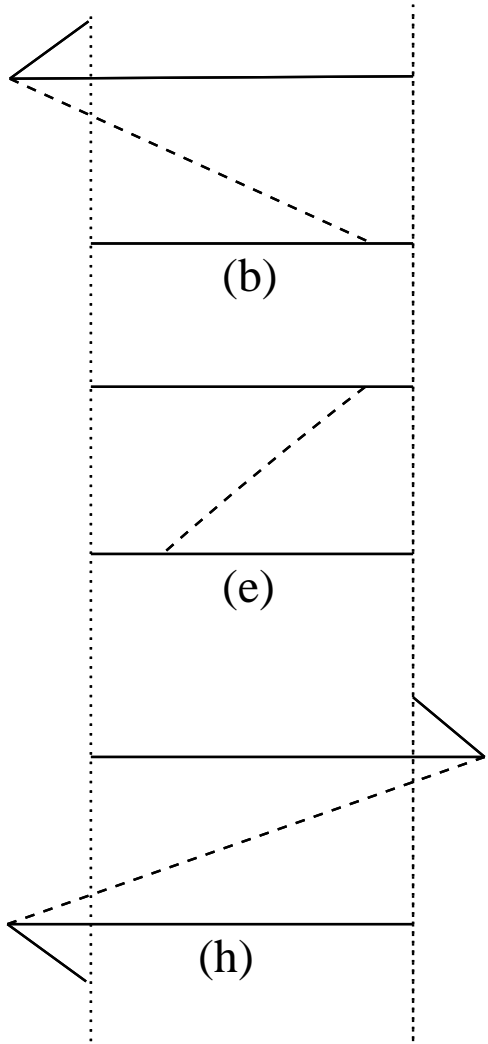
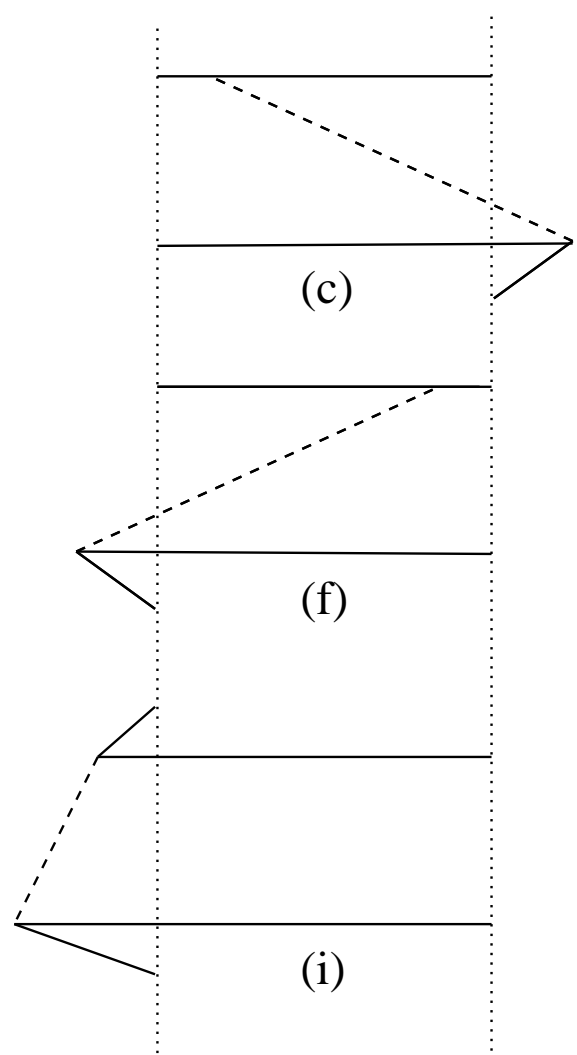
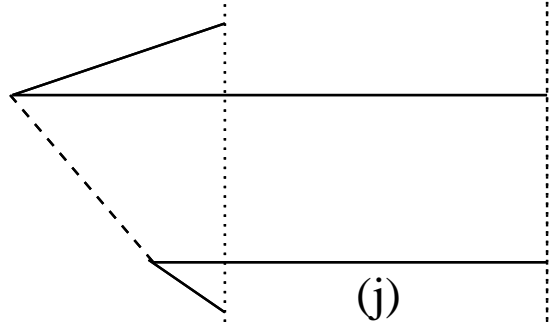
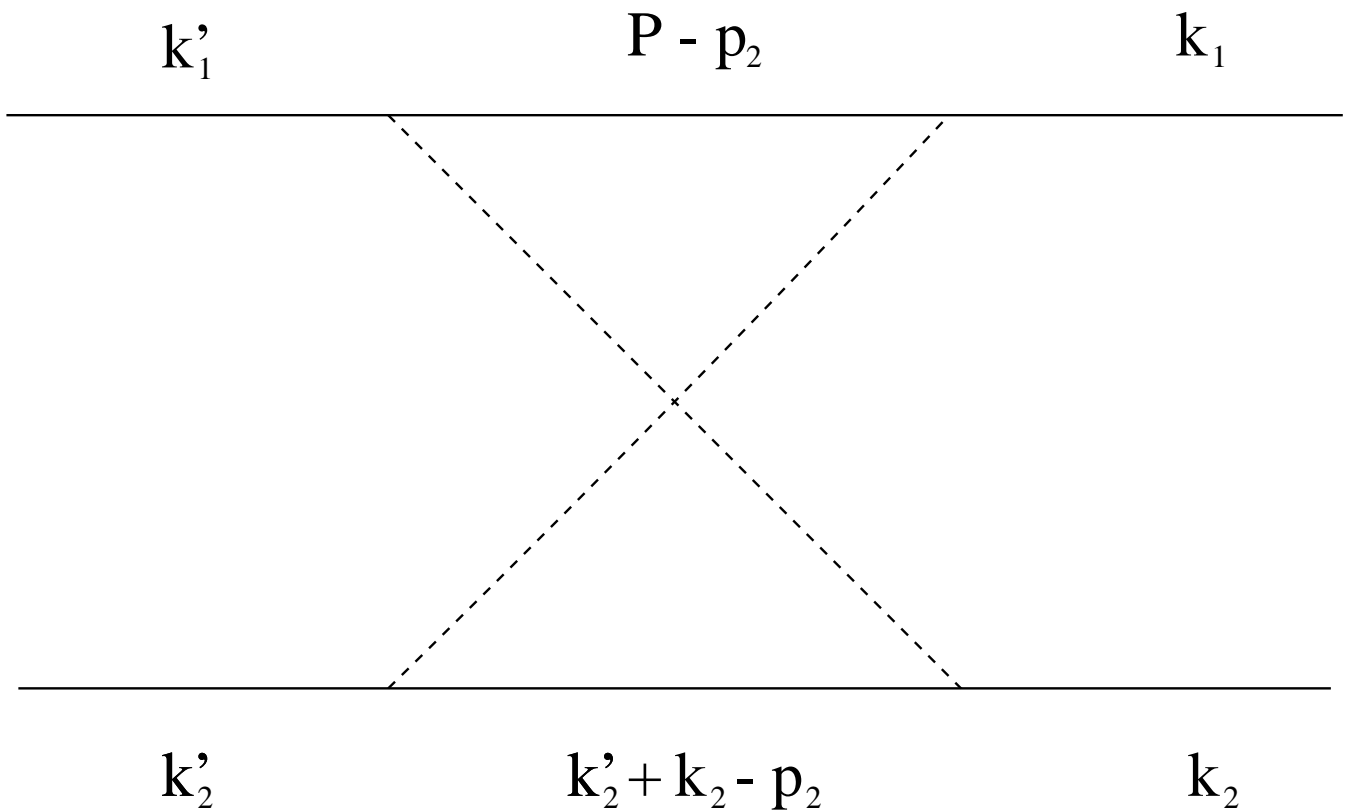
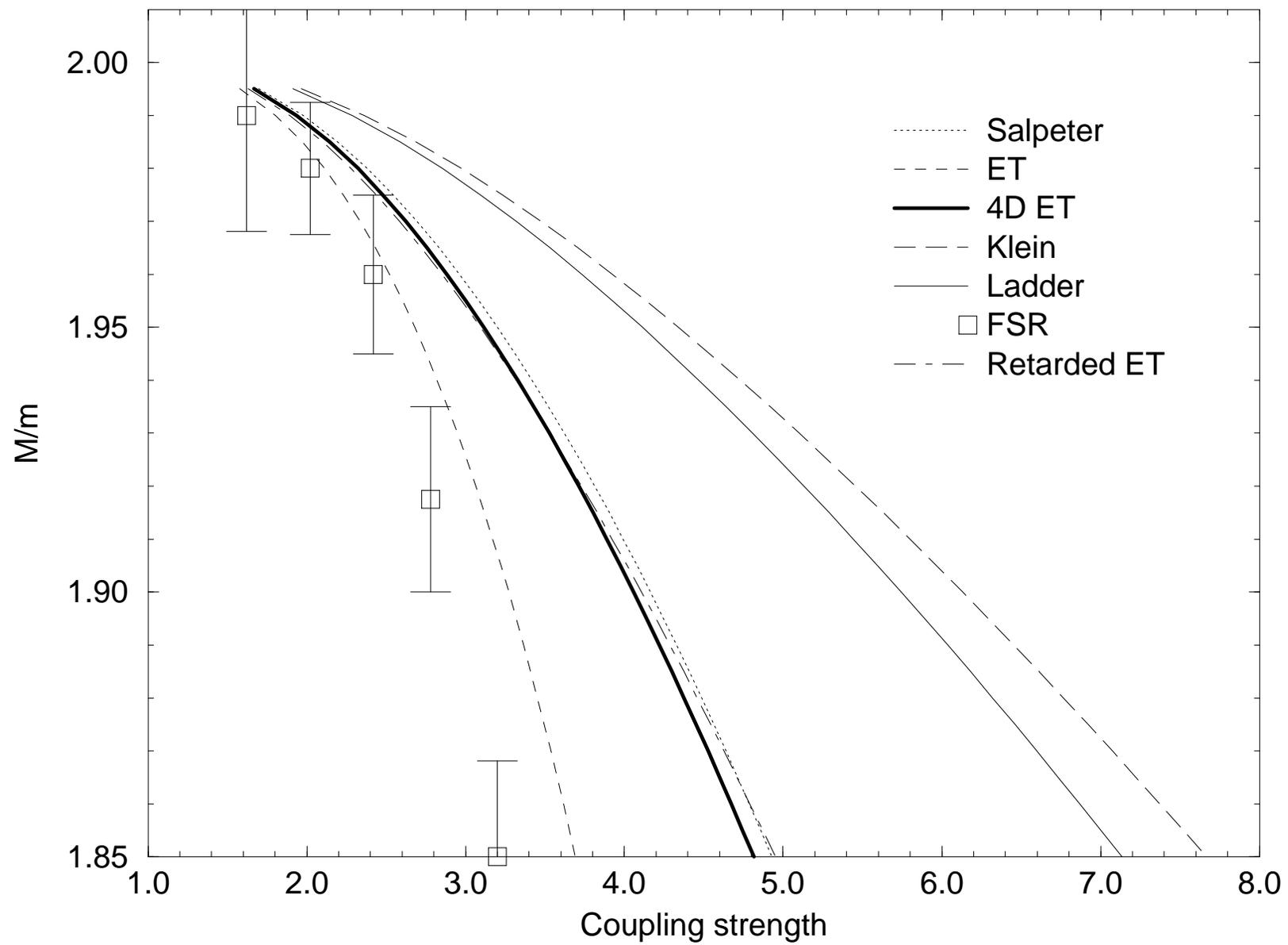
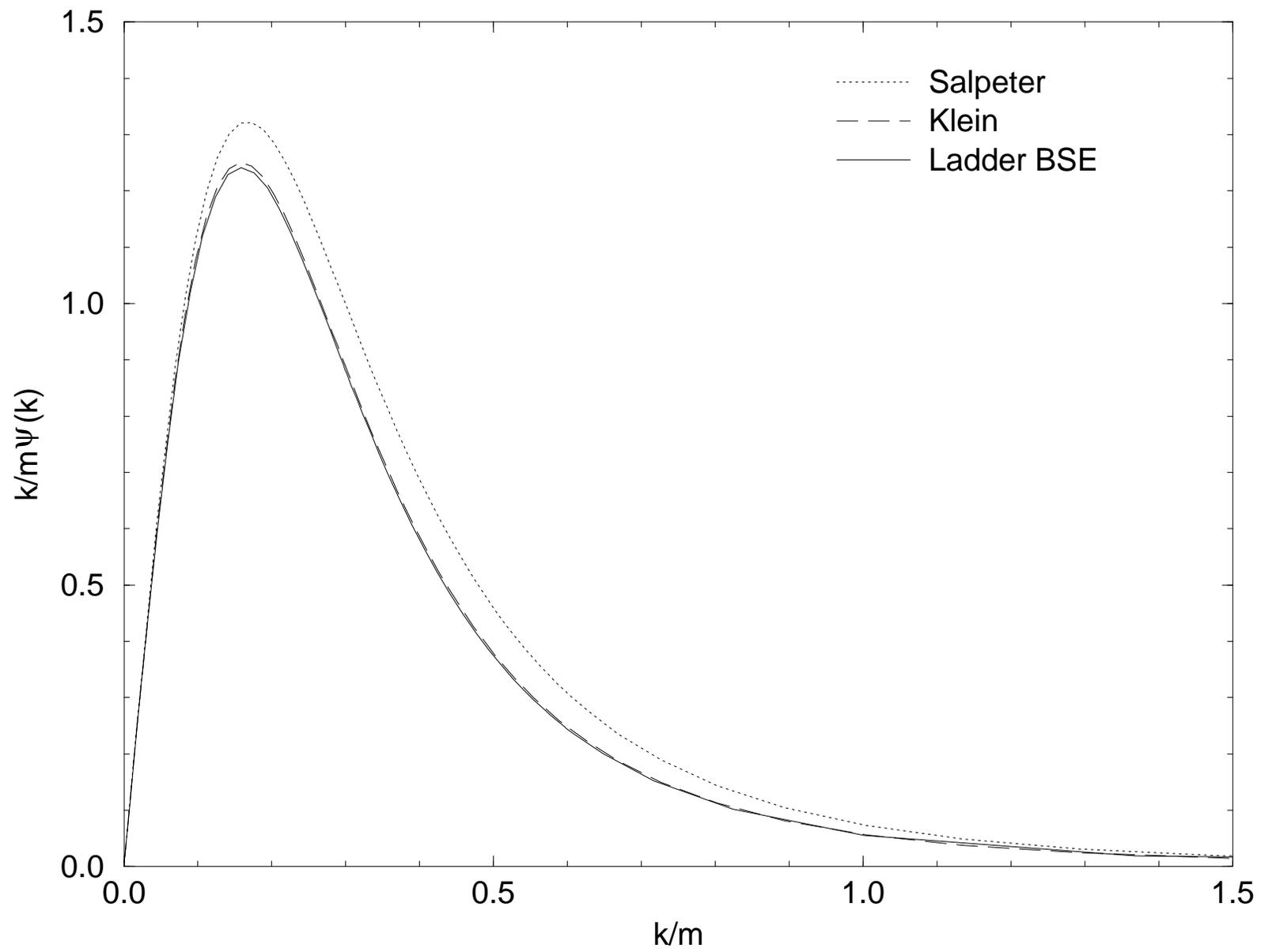
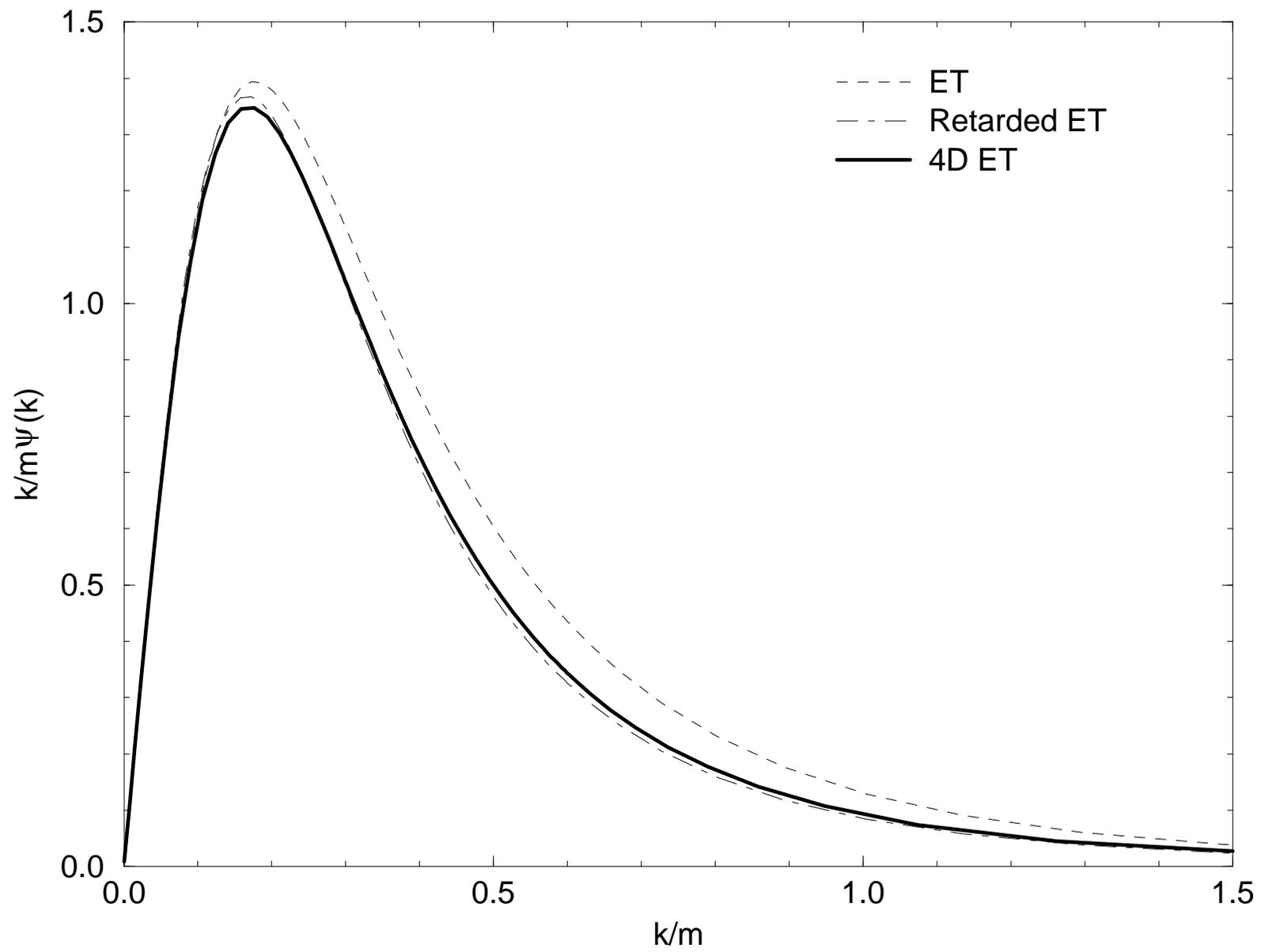
\end{document}